\documentclass[journal]{IEEEtran}
\usepackage{amsmath,amsfonts}
\usepackage{algorithmic}
\usepackage{algorithm}
\usepackage{array}
\usepackage{textcomp}
\usepackage{booktabs}
\usepackage{tabularx}
\usepackage{array}
\usepackage{amsmath}
\usepackage{url}
\usepackage{verbatim}
\usepackage{graphicx}
\usepackage{amssymb} 
\usepackage{amsfonts}
\usepackage{amsmath} 
\usepackage{physics} 
\usepackage{algorithmic}  
\usepackage{algorithm} 
\usepackage{graphicx}
\usepackage{color} 
\usepackage{xcolor}
\usepackage{bm}
\usepackage{subfigure}
\usepackage{booktabs}  
\usepackage[dvipsnames]{xcolor} 
\usepackage{cite} 
\usepackage{balance}
\usepackage{setspace}  
\usepackage{amsmath} 
\usepackage[colorlinks=true,linkcolor=blue,citecolor=blue,urlcolor=blue,breaklinks=true,pdfborder={0 0 0}]{hyperref}
\usepackage{float}
\newcounter{theorem}[section]
\newcounter{proposition}[section]
\newcounter{lemma}[section]
\newcounter{remark}[section]
\newcounter{definition}[section]

\def\ththeorem{\arabic{section}.\arabic{theorem}}

\def\thelemma{\arabic{section}.\arabic{lemma}}
\def\theremark{\arabic{section}.\arabic{remark}}
\def\thedefinition{\arabic{section}.\arabic{definition}}

\newcommand{\algref}[1]{\textbf{Algorithm \ref{#1}}}
\newcommand{\tabref}[1]{\textbf{Table \ref{#1}}}

\newcommand{\figref}[1]{Fig. \ref{#1}}
\newcommand{\secref}[1]{Section \ref{#1}}
\DeclareMathAlphabet{\mathsfit}{\encodingdefault}{\sfdefault}{m}{sl}
\SetMathAlphabet{\mathsfit}{bold}{\encodingdefault}{\sfdefault}{bx}{n}
\newcommand{\tens}[1]{\bm{\mathsfit{#1}}}
\def\tA{{\tens{A}}}
\def\tB{{\tens{B}}}

\def\tC{{\tens{C}}}

\def\tE{{\tens{E}}}
\def\tF{{\tens{F}}}

\def\tH{{\tens{H}}}

\def\tS{{\tens{S}}}
\def\tT{{\tens{T}}}

\def\tX{{\tens{X}}}

\begin{document}
	
\title{Robust Decentralized Multi-Satellite Massive MIMO Transmission via Knowledge Distillation}


	\author{Wenjing Cao, \textit{Graduate Student Member}, \textit{IEEE},
    Zheng Lin, \textit{Member}, \textit{IEEE}, \\
    Yafei Wang, \textit{Graduate Student Member}, \textit{IEEE},  Wenjin Wang, \textit{Member}, \textit{IEEE}, Ye Wang, \textit{Member}, \textit{IEEE},\\
   Rui Ding, Symeon Chatzinotas, \textit{Fellow}, \textit{IEEE}, Bj{\"o}rn Ottersten, \textit{Fellow}, \textit{IEEE}
		\thanks{Manuscript received xxx; revised xxx.  \textit{}}
		\thanks{Wenjing Cao, Yafei Wang, and Wenjin Wang are with the National Mobile Communications Research Laboratory, Southeast University, Nanjing 210096, China, and also with Purple Mountain Laboratories, Nanjing 211100, China (e-mail: caowj@seu.edu.cn; wangyf@seu.edu.cn; wangwj@seu.edu.cn).}
	    \thanks{Zheng Lin, Symeon Chatzinotas, and Bj{\"o}rn Ottersten are with the Interdisciplinary Centre for Security, Reliability and Trust (SnT), University of Luxembourg, Luxembourg (e-mail: zhenglin@ieee.org; symeon.chatzinotas@uni.lu; bjorn.ottersten@uni.lu).}
        \thanks{Ye Wang is with Peng Cheng Laboratory, Shenzhen 518055, China (e-mail: wangy02@pcl.ac.cn). Rui Ding is with China Satellite Network Group Company Ltd., Beijing 100029, China (e-mail: greatdn@qq.com)}}

\markboth{}%
{Cao \MakeLowercase{\textit{et al.}}: Robust Decentralized Multi-Satellite Massive MIMO Transmission via Knowledge Distillation}
	
\maketitle
\begin{abstract}
This paper investigates robust decentralized transmission for cooperative multi-satellite massive multiple-input multiple-output (MIMO) systems under imperfect statistical channel state information (sCSI). 
In the considered scenario, each satellite has complete access to its local information but receives partial information from other satellites due to limited inter-satellite links (ISLs), with only imperfect sCSI available.
\textcolor{black}{To address these challenges, we propose a knowledge distillation (KD) framework that transfers cooperative precoding knowledge from a centralized teacher neural network (NN) to lightweight decentralized student NNs. Specifically, a \textit{global-clean} teacher, aggregating information from all satellites and accessing accurate sCSI during offline training, transfers its cooperative precoding knowledge to \textit{partial-noisy} students, relying on complete local information, limited information exchanged by other satellites, and error-corrupted sCSI for local precoding.} The teacher NN combines patch-wise self-attention with dual-axis attention to learn inter-user interference and inter-satellite coordination, whereas each student NN adopts a compact per-satellite architecture for efficient onboard inference. The teacher learns a high-quality weighted minimum mean square error precoding policy from global-clean inputs, which is then distilled into the students operating on partial-noisy inputs. To mitigate the resulting teacher-student performance gap, we develop a hybrid KD mechanism with explicit angle- and phase-error calibration. 
Simulation results demonstrate that the proposed framework significantly enhances the decentralized sum-rate performance and remains robust under diverse configurations.
\end{abstract}

	%
	\IEEEpeerreviewmaketitle
\vspace{-2mm}
\section{Introduction}
\vspace{-1mm}
\IEEEPARstart{S}{atellite} networks are expected to play a pivotal role in future wireless systems by extending ubiquitous connectivity beyond the reach of terrestrial infrastructures \cite{WANG202542, Pan2026, Bakhsh2025}. In particular, dense low-earth-orbit (LEO) constellations have garnered significant attention owing to their low propagation delay, wide-area coverage, and deployment flexibility \cite{Shang2026, cao2025}. Nevertheless, single-satellite transmission faces fundamental bottlenecks, including severe path loss, limited service capacity, and intense inter-user interference, especially under full frequency reuse and massive user access \cite{WANG_2026MSCT}. Cooperative multi-satellite massive multiple-input multiple-output (MIMO) transmission provides a promising means of overcoming these limitations by jointly exploiting the large-scale antenna resources distributed across multiple satellites \textcolor{black}{that share a common coverage area}\cite{Bakhsh2025, wang2026SemMSMS}. 
Specifically, the large-scale antenna array onboard each satellite provides substantial beamforming gains to mitigate severe propagation loss, while coordinated multi-satellite transmission aggregates distributed antenna resources to provide additional spatial degrees of freedom for multi-user and multi-stream transmission. 
These spatial degrees of freedom also enable more flexible precoding for inter-user interference suppression under aggressive frequency reuse, thereby improving service capacity and spectral efficiency. Efficient precoding is therefore essential to fully exploit these spatial-domain benefits in cooperative multi-satellite massive MIMO systems \cite{Xiang2024,cao2026deep}.

In cooperative multi-satellite massive MIMO systems, precoding enables multiple satellites to jointly transmit over shared time-frequency resources while mitigating inter-user and inter-satellite interference. Conventional centralized schemes \textcolor{black}{assume that a network} controller collects global channel state information (CSI) to jointly optimize precoders of all satellites, typically solving the weighted sum-rate (WSR) maximization problem via iterative algorithms \cite{Xiang2024}. Recent learning-based approaches utilize neural networks (NNs) trained on global statistical CSI (sCSI) for rapid cooperative precoding inference \cite{wang2026, wang2025MSMS}. Despite the spectral-efficiency advantages of centralized multi-satellite cooperation, they rely on globally available CSI, substantial computational power, and frequent inter-satellite signaling. These requirements are unsustainable in large-scale LEO constellations characterized by limited onboard resources and bandwidth-constrained, topology-varying inter-satellite links (ISLs) \cite{Rodrigues2023, Chu2025, Beatriz2019}. Consequently, decentralized multi-satellite transmission has emerged as a promising alternative, wherein each satellite infers its local precoder using only local sCSI and restricted ISL-exchanged data \cite{cao2026deep}. Nevertheless, there are two critical challenges for deploying decentralized multi-satellite transmission. First, while most existing designs predicate their performance on accurate sCSI, practical LEO links suffer from severe sCSI impairments—such as angle-estimation errors, attitude jitter, array calibration errors and uncertainty, and feedback delays—making them highly sensitive to angle-of-arrival (AoA) and angle-of-departure (AoD) inaccuracies \cite{You2020, Cilden2019, pan2019, Liu2022}. Second, decentralized learning-based schemes operate with noisy partial sCSI under a lightweight NN, resulting in a significant performance bottleneck compared with centralized schemes that rely on clean global sCSI.

Knowledge distillation (KD) has been widely studied as an effective technique for enhancing deployment efficiency, achieving broad success across computer vision, natural language processing, and foundation models \cite{hinton2015, gou2021kd, Hu2023kd, Amir2025kd}. While classical KD transfers softened output distributions from a high-capacity teacher to a compact student \cite{hinton2015}, subsequent advancements have shifted the distillation target to intermediate representations \cite{gou2021kd}. 
Beyond conventional model compression, clean-to-noisy KD offers a compelling paradigm for robust wireless transmission \cite{Hong2021}. Specifically, \cite{Hong2021} demonstrates that a teacher trained on pristine samples can provide hidden-feature supervision to a student processing corrupted inputs, guiding the student to learn denoised, generalizable representations. 
In the physical layer, KD has proven effective across various representative tasks. For CSI feedback, high-capacity autoencoders transfer reconstruction capabilities to lightweight student NNs \cite{Cui2024CSIKD}. 
Similarly, in beam prediction, KD facilitates compact mapping models that approach teacher-level accuracy and spectral efficiency at a fraction of the computational cost \cite{Tavakolian2026BeamKD}. 
Furthermore, KD can compress and exchange low-dimensional representations among cooperative agents, significantly curbing inter-agent communication overhead while retaining task-aware domain knowledge \cite{Liu2026}. Collectively, these findings validate that KD can effectively transfer coordination-aware and noise-robust policies from an offline centralized teacher (operating on clean global sCSI) to decentralized students (constrained by noisy partial sCSI). This mechanism elegantly bridges the gap between idealized centralized design and practical decentralized implementation.

From a KD perspective, although centralized cooperative schemes incur substantial signaling and computational overheads that preclude practical deployment, they can serve as highly effective teachers during offline training. By leveraging clean global sCSI, they extract near-optimal precoding policies and provide high-quality supervision targets, thereby circumventing the need for computationally intensive alternating iterations during inference \cite{wang2026,wang2025MSMS, Xiang2024}. Conversely, decentralized schemes offer a lightweight and efficient onboard inference framework. However, their reliance on noisy, partial sCSI and limited ISL data corrupted by angle errors fundamentally restricts their capacity to acquire global cooperative knowledge \cite{Liu2022, cao2026deep}. This inherent asymmetry aligns well with the KD paradigm, wherein a high-capacity teacher with access to pristine global information transfers task-relevant representations to lightweight, decentralized students operating under imperfect observations \cite{Hong2021}. Motivated by these observations, this paper addresses the following fundamental research question: \textit{\textcolor{black}{How to design efficient and robust decentralized multi-satellite cooperative transmission under partial and imperfect sCSI?}} To resolve this challenge, our principal contributions are summarized as follows:
\begin{itemize}  
\item We propose a KD learning framework for cooperative multi-satellite transmission, where a large centralized teacher NN is developed to guide decentralized student NNs. The teacher NN mainly consists of a patch-wise self-attention encoder and a dual-axis attention module: the former refines feature representations within each satellite-user pair, while the latter captures user-domain interference relationships and satellite-domain cooperative dependencies through user-axis and satellite-axis attention, respectively. 
By leveraging clean and global CSI inputs, the centralized teacher NN learns a high-performance precoding policy that approaches the performance of WSR maximization, thereby serving as reliable and efficient supervision targets for decentralized student NNs with low inference complexity, avoiding the need to obtain targets through computationally intensive alternating iterations.
\item We develop lightweight decentralized student NNs that operate with practical partial and angle-error-corrupted observations at individual satellites. Specifically, each student replaces the deep patch-wise encoder with a compact single-layer counterpart, and infers only its local precoder, which facilitates effective knowledge transfer during distillation while reducing computational complexity for onboard inference. Unlike the centralized teacher, each student relies only on locally available sCSI and a small amount of ISL-exchanged information, both of which are corrupted by angle errors. 

\item We design an angle- and phase-correction-based hybrid KD mechanism to bridge the gap between the centralized teacher and decentralized students. The proposed mechanism guides each student NN to predict angle residuals and phase residuals during training. The corrected angle information is then converted into Kronecker-structured steering vectors and combined with phase correction to obtain more accurate array responses for robust decentralized closed-form precoding. In addition, hybrid distillation transfers physically interpretable knowledge, including corrected steering vectors and precoders, from the teacher to the students. Simulation results demonstrate that the proposed mechanism improves the sum-rate performance of decentralized student NNs under partial and noisy inputs, while maintaining robustness and generalization across different angle-error distributions and satellite-user scales.
\end{itemize}

This paper is structured as follows: \secref{System} introduces the multi-satellite system model and problem formulation. \secref{Multi-Satellite_Cooperation} presents the large centralized teacher NN and lightweight decentralized student NNs design. \secref{Knowledge_Distillation_Design} presents the KD design. \secref{simulation} provides the simulation results. \secref{conclusion} concludes this article.

{\textit{Notation}}: $x, {\bf x}, {\bf X}$ represent scalar, column vector, matrix. $(\cdot)^T$, $(\cdot)^{*}$, $(\cdot)^H$, and $(\cdot)^{-1}$ respectively denote the transpose, conjugate, transpose-conjugate, and inverse operations. $\left \|\cdot\right \|_{2}$ denotes $l_2$-norm. $\otimes$ and $\odot$ are the Kronecker product and Hadamard product operations. The operator ${\rm Tr}\left\{\cdot\right\}$ represents the matrix trace. $\mathbb{E}\left\{\cdot\right\}$ denotes the expectation. The expression $\mathcal{C}\mathcal{N}(\mu, \sigma^2)$ denotes circularly symmetric complex  Gaussian distribution with expectation $\mu$ and variance $\sigma^2$. ${\mathbb{R}}^{M\times N}$ and ${\mathbb{C}}^{M\times N}$ represent the set of $M\times N$ dimension real- and complex-valued matrices.  $k\in \mathcal{K}$ means element $k$ belongs to set $\mathcal{K}$. $\Re(\cdot) $ and $\Im(\cdot)$ denote the real and imaginary parts of a complex scalar, vector, or matrix.
We use  $\tX_{[m_{1}, \ldots, m_{N}]}$ to denote the indexing of elements in tensor  $\tX \in \mathbb{R}^{M_{1} \times \cdots \times M_{N}} .[\tA_{1}, \ldots, \tA_{K}]_{S}$  denotes the tensor formed by stacking  $\tA_{1}, \ldots, \tA_{K}$  along the  $S$-th dimension.  $[\cdot]_{0}$ denotes the concatenation of tensors along a new dimension, i.e.,  $\tA_{[n,:, \cdots]}=\tB_{n}, \forall n$  when $\tA=\left[\tB_{1}, \ldots, \tB_{N}\right]_{0}$. We define the product of tensor $\tX \in \mathbb{R}^{M_{1} \times \cdots \times M_{N} \times D_{X}}$ and matrix $\mathbf{Y} \in \mathbb{R}^{D_{X} \times D_{Y}}$  as  $(\tX \times \mathbf{Y})_{[m_{1}, \ldots, m_{N},:]}= \tX_{[m_{1}, \ldots, m_{N},:]} \mathbf{Y}$.
\section{Multi-Satellite Transmission Model and WMMSE Closed-Form Structure}
\label{System}
\begin{figure}[!t]
	\centering
\includegraphics[width=0.75\linewidth,trim=0.8cm 0.52cm 0.3cm 0.3cm,clip]{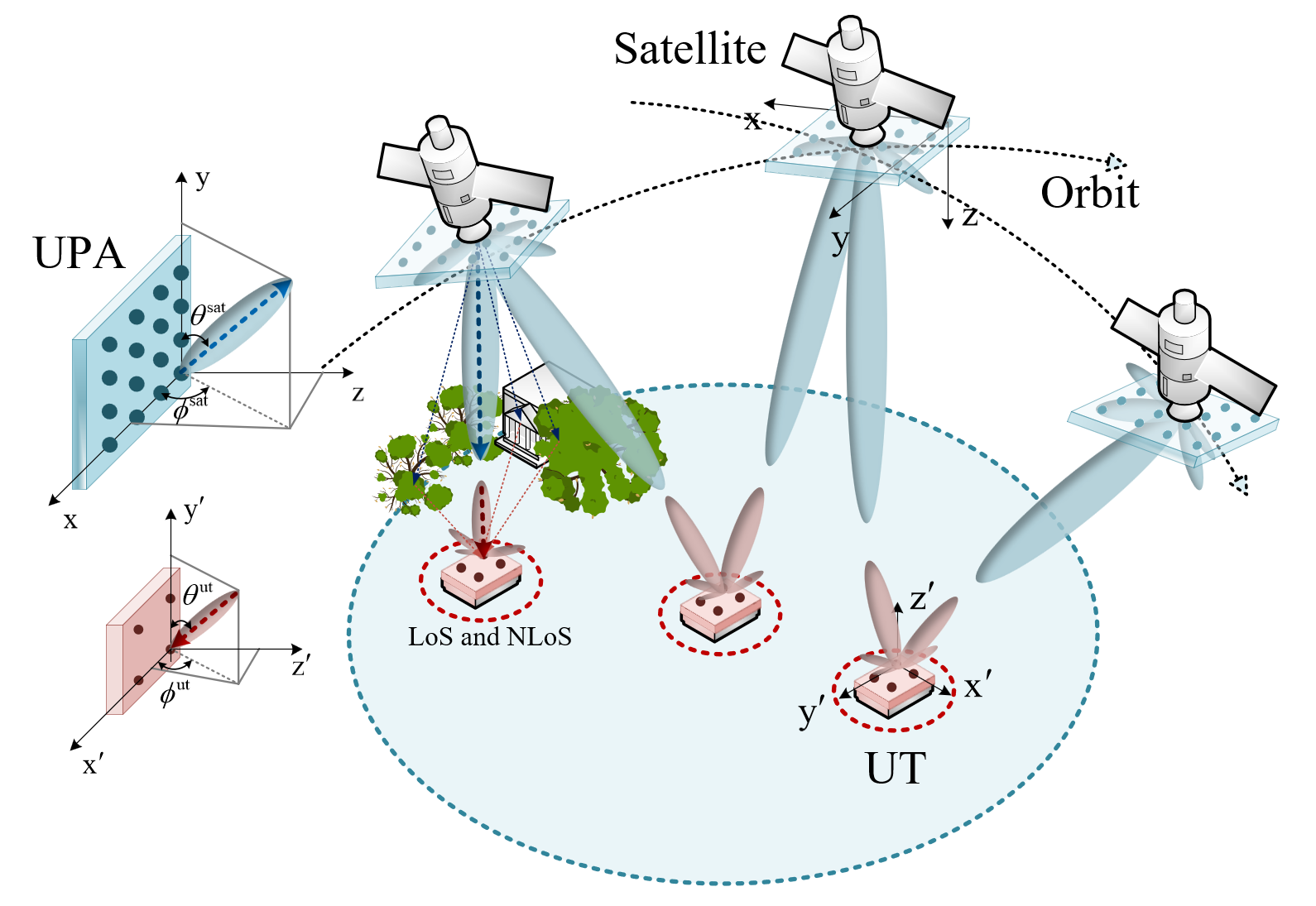}
\caption{System architecture of cooperative multi-satellite transmission.}
\label{MSIT_scenario}
\vspace{-5mm}
\end{figure}

We consider a downlink (DL) multi-satellite cooperative transmission system, as shown in \figref{MSIT_scenario}, comprising $S$ LEO satellites, performing transmission to $K$ \textcolor{black}{mobile} user terminals (UTs) within the same time-frequency resources. For clarity, we define the satellite set as $\mathcal{S} = \{1, \ldots, S\}$ and the UT set as $\mathcal{K} = \{1, \ldots, K\}$. Each satellite is equipped with a large-scale uniform planar array (UPA) consisting of  $M = M_{\text{x}} \times M_{\text{y}}$  antenna elements, where  $M_{\text{x}}$  and  $M_{\text{y}}$  represent the number of antennas along the x-axis and y-axis, respectively. Similarly, each UT is equipped with a UPA composed of \(N = N_{\mathrm{x}^{\prime}} \times N_{\mathrm{y}^{\prime}}\) antenna elements.
In cooperative multi-satellite multi-stream (MSMS) transmission, non-coherent transmission enables multiple satellites to send independent data streams to the same UT equipped with an antenna array, while all beams simultaneously reuse the full system bandwidth and time-frequency resources \cite{Shang2026,wang2026}. 
Unlike coherent cooperation, which requires stringent synchronization \cite{zhu2026, Marrero2022}, the considered non-coherent scheme offers greater implementation flexibility while improving spectral efficiency through spatial multiplexing \cite{WANG202542}.

\vspace{-3mm}
\subsection{Channel and Signal Models}
\vspace{-1mm}

To support broadband communication in the multi-satellite system, the entire system bandwidth is employed for the orthogonal frequency division multiplexing (OFDM) transmission. Let the number of subcarriers be denoted by \(N_{\rm sc}\), the cyclic prefix (CP) length by \(N_{\rm cp}\), and the system sampling period by \(T_{\rm sam}\). The CP duration is given by \(T_{\rm cp}=N_{\rm cp}T_{\rm sam}\). The OFDM symbol duration without CP is \(T_{\rm sc}=N_{\rm sc}T_{\rm sam}\), while the total symbol duration including CP is \(T_{\rm sym}=T_{\rm sc}+T_{\rm cp}\). The time-domain transmit signal of the \(s\)-th satellite during the \(m\)-th OFDM symbol is given by  \(\mathbf{x}_{s,m}(t)\in\mathbb{C}^{M\times1}\):
\begin{align}
\!\!\!\!\mathbf{x}_{s,m}(t)\!\textstyle=\!\!\! \sum_{r=0}^{N_{\rm sc}-1} \!\!\mathbf{x}_{s,m,r} {\mathrm e}^{j2\pi r \Delta f t},  - T_{\rm cp} \!\leq\! t\!-\!mT_{\rm sym}\! < \!T_{\rm sc},
\end{align}
herein, $\mathbf{x}_{s,m,r}$ denotes the frequency-domain transmit signal vector over the $r$-th subcarrier in the $m$-th OFDM symbol and $\Delta f =1/T_{\rm sc}$. The DL time-domain received signal vector of the $k$-th UT from the $s$-th satellite in the $m$-th OFDM symbol at the $t$-th time instant can be written as:
\begin{align}
\mathbf{y}_{s,k,m}(t)\textstyle=\int_{-\infty}^{\infty} \tilde{\mathbf{H}}_{s,k}(t,\tau)\mathbf{x}_{s,m}(t-\tau){\rm d}\tau+\mathbf{z}_{s,k,m}(t),
\end{align}
\textcolor{black}{where we denote $\tau$ as the delay variable,} $\mathbf{z}_{s,k,m}(t)=\sum_{s^{\prime} \in \mathcal{S} \setminus \{s\}} \int_{-\infty}^{\infty} \tilde{\mathbf{H}}_{s^{\prime},k}(t, \tau) \mathbf{x}_{s^{\prime},m}(t-\tau){\rm d}\tau+ \tilde{\mathbf{n}}_{k}(t)$, and $\tilde{\mathbf{n}}_{k}(t) \in \mathbb{C}^{N \times 1}$ denotes
the additive white Gaussian noise vector with distribution $\mathcal{C N}(\mathbf{0}, \sigma_{k}^{2} \mathbf{I}_N)$.
Herein, the time-varying spatial
domain MIMO channel $\tilde{\mathbf{H}}_{s,k}(t,\tau) \in \mathbb{C}^{N \times M}$ from the $s$-th satellite to the $k$-th UT can be expressed as \cite{Li2022}:
\begin{align}
\!\!\!\!\tilde{\mathbf{H}}_{s,k}(t,\tau)\!\textstyle=\!\!\!\sum_{\ell=0}^{L_{s,k}\!-\!1} \!\!\!\!\!a_{s,k, \ell} {\mathrm e}^{j 2 \pi \nu_{s,k,\ell} t} \delta(\tau\!-\!\tau_{s,k,\ell})\mathbf{d}_{s,k, \ell}\mathbf{g}_{s,k,\ell}^H,
\end{align}
where $L_{s,k}$ is the multipath number of the channel from the $s$-th satellite to the $k$-th UT, $a_{s,k,\ell}$, $\nu_{s,k,\ell}$, $\tau_{s,k,\ell}$, $\mathbf{d}_{s,k,\ell}$ and $\mathbf{g}_{s,k,\ell}$ are the complex channel gain, Doppler shift, propagation delay, receive array response vector and transmit array response vector associated with the $\ell$-th path.
Specifically, the AoD at the satellite side is denoted by $(\theta_{s,k,\ell}^{\rm sat}, \phi_{s,k,\ell}^{\rm sat})$, the AoA at the UT side is denoted by $(\theta_{s,k,\ell}^{\rm ut}, \phi_{s,k,\ell}^{\rm ut})$.
In LEO satellite channels, owing to the large satellite-to-ground distance and the fact that scattering occurs within a few kilometers around UTs, the AoD pairs corresponding to different multipath components of the same link are nearly identical. Therefore, we have
$(\theta_{s,k,\ell}^{\rm sat}, \phi_{s,k,\ell}^{\rm sat})\! =\! (\theta_{s,k}^{\rm sat}, \phi_{s,k}^{\rm sat}), \forall \ell$.
The transmit array response vectors at the satellite side are \cite{You2020, Liu2022}:
\begin{align}
\mathbf{g}_{s,k,\ell}\! =\! \mathbf{g}_{s,k}\! =\! \mathbf{a}_{M_{\rm x}}\!\!\left(\sin \theta^{\rm sat}_{s,k} \cos \phi^{\rm sat}_{s,k}\right)\! \otimes \!\mathbf{a}_{M_{\rm y}}\!\!\left(\cos \theta^{\rm sat}_{s,k}\right).
\end{align}
The receive array response vectors are given by:
\begin{align}
\mathbf{d}_{s,k,\ell} = \mathbf{a}_{N_{\rm x^{\prime}}}\left(\sin \theta^{\rm ut}_{s,k,\ell} \cos \phi^{\rm ut}_{s,k,\ell}\right) \otimes \mathbf{a}_{N_{\rm y^{\prime}}}\left(\cos \theta^{\rm ut}_{s,k,\ell}\right).
\end{align}
We define $\mathbf{a}_{n_{\rm v}}(x) = \frac{1}{\sqrt{n_{\rm v}}}[1, \mathrm{e}^{-\frac{j 2 \pi}{\lambda}d_{\rm v}x}, \ldots, \mathrm{e}^{-\frac{j 2 \pi}{\lambda} d_{\rm v}(n_{\rm v}-1)x}]^{T}$. \( d_{\rm v} = \frac{\lambda}{2} \) represents the distance between adjacent antenna elements along the $\rm v$-axis and \( {\rm v}\in \left\{\rm x,\rm y,\rm x^{\prime},\rm y^{\prime} \right\} \).
To separate the signals from multiple satellites at the multi-antenna UT, spatial linear receive processing is applied as:
\begin{align}
&r_{s,k,m}(t)=\mathbf{b}_{s,k}^{H} \mathbf{y}_{s,k,m}(t)\nonumber\\
&\textstyle=\mathbf{b}_{s,k}^{H} \left(\int_{-\infty}^{\infty} \tilde{\mathbf{H}}_{s,k}(t,\tau)\mathbf{x}_{s,m}(t-\tau){\rm d}\tau+\mathbf{z}_{s,k,m}(t)\right),
\end{align}
\( \mathbf{b}_{s,k} \!\!\in\!\! \mathbb{C}^{N \times 1}\) denotes the receive vector at the $k$-th UT for extracting the signal from the $s$-th satellite. \textcolor{black}{To compensate for the large and predictable Doppler shift and propagation delay in LEO satellite links, a two-stage compensation procedure is considered. First, each satellite performs bulk Doppler and delay pre-compensation toward the beam center based on its ephemeris and beam-pointing information. The $k$-th UT then estimates the remaining Doppler shift and propagation delay of each satellite primarily from the received DL pilots, with ephemeris and global navigation satellite system (GNSS) information providing auxiliary position-based post-compensation \cite{Wang2019, Xiang2024}.  Let $\nu_{s,k}^{\rm sat}$ and $\tau_{s,k}^{\rm sat}$ denote the satellite-side bulk pre-compensation terms, while $\nu_{s,k}^{\rm cps}$ and $\tau_{s,k}^{\rm cps}$ model the remaining Doppler and delay errors after beam-center-based compensation at the $k$-th UT \cite{zhu2026}, respectively. After compensation, the compensated signal at the $k$-th UT is:}
\begin{align}
r_{s,k,m}^{\rm cps}(t)
=&
r_{s,k,m}(t+\tau_{s,k}^{\rm cps})
{\rm e}^{-j2\pi\nu_{s,k}^{\rm cps}(t+\tau_{s,k}^{\rm cps})}
\nonumber\\
=&
\mathbf b_{s,k}^{H}
\Big(\textstyle\sum_{\ell=0}^{L_{s,k}-1}
a_{s,k,\ell}
{\rm e}^{j2\pi\nu_{s,k,\ell}^{\rm ut}
(t+\tau_{s,k}^{\rm cps})}\nonumber\\&
\mathbf d_{s,k,\ell}\mathbf g_{s,k}^{H}
\mathbf x_{s,m}
\left(t-\tau_{s,k,\ell}^{\rm ut}\right)\Big)
+
z_{s,k,m}^{\rm cps}(t),
\end{align}
$\nu_{s,k,\ell}$ and $\tau_{s,k,\ell}$ denote the original Doppler shift and propagation delay of the $(s,k,\ell)$-th path in the channel model. The residual Doppler shift and delay after the satellite-side pre-compensation and UT-side post-compensation are defined as
$\nu_{s,k,\ell}^{\rm ut}
\triangleq
\nu_{s,k,\ell}
-\nu_{s,k}^{\rm sat}
-\nu_{s,k}^{\rm cps},
\tau_{s,k,\ell}^{\rm ut}
\triangleq
\tau_{s,k,\ell}
-\tau_{s,k}^{\rm sat}
-\tau_{s,k}^{\rm cps},
$ respectively. Since the compensation is performed with respect to the intended satellite $s$, the desired satellite-UT link is approximately synchronized in time and frequency after compensation, whereas signals from other satellites remain asynchronous. $z_{s,k,m}^{\rm cps}(t)$ denotes the aggregate inter-satellite interference-plus-noise term. Assume that the residual delay spread after compensation is within the CP
duration, i.e.,
$\max_{\ell}\tau_{s,k,\ell}^{\rm ut}\leq T_{\rm cp}$. Then, over the useful
OFDM symbol interval, $\mathbf x_{s,m}
(t-\tau_{s,k,\ell}^{\rm ut})
=
\sum_{q=0}^{N_{\rm sc}-1}
\mathbf x_{s,m,q}
{\rm e}^{j2\pi q\Delta f
(t-\tau_{s,k,\ell}^{\rm ut})}$.
The effective MIMO channel frequency response is:
\begin{align}
\mathbf H_{s,k}(t,f)
=
\mathbf d_{s,k}(t,f)\mathbf g_{s,k}^{H}
\in\mathbb C^{N\times M},
\end{align}
where $\mathbf d_{s,k}(t,f)
=
\sum_{\ell=0}^{L_{s,k}-1}
\bar a_{s,k,\ell}
{\rm e}^{j2\pi
\left(
\nu_{s,k,\ell}^{\rm ut}t
-
f\tau_{s,k,\ell}^{\rm ut}
\right)}
\mathbf d_{s,k,\ell}$ and $\bar a_{s,k,\ell}
\triangleq
a_{s,k,\ell}
{\rm e}^{j2\pi\nu_{s,k,\ell}^{\rm ut}\tau_{s,k}^{\rm cps}}$. Therefore, the compensated time-domain
signal can be rewritten as:
$r_{s,k,m}^{\rm cps}(t)
=
\mathbf b_{s,k}^{H}\sum_{q=0}^{N_{\rm sc}-1}
\mathbf H_{s,k}(t,q\Delta f)
\mathbf x_{s,m,q}
{\rm e}^{j2\pi q\Delta f t}
\!+\!
z_{s,k,m}^{\rm cps}(t)$.
After removing the CP, the received signal over the $r$-th subcarrier is
obtained by OFDM demodulation as:
\begin{align}
\!\!\!\!&r_{s,k,m,r}
\textstyle=
\frac{1}{T_{\rm sc}}
\int_{mT_{\rm sym}}^{mT_{\rm sym}+T_{\rm sc}}
r_{s,k,m}^{\rm cps}(t)
{\rm e}^{-j2\pi r\Delta f t}
{\rm d}t
\\
&\textstyle=\!\!
\frac{1}{T_{\rm sc}}\!\!\!
\sum_{q=0}^{N_{\rm sc}-1}\!\!\!
\int\!\mathbf b_{s,k}^{H}
\mathbf H_{s,k}(t,q\Delta f)
\mathbf x_{s,m,q} {\rm e}^{j2\pi(q-r)\Delta f t}
{\rm d}t
\!+\!
z_{s,k,m,r},\nonumber
\end{align}
$z_{s,k,m,r}
\!\!=\!\!
\frac{1}{T_{\rm sc}}\!\!\!
\int_{mT_{\rm sym}}^{mT_{\rm sym}+T_{\rm sc}}\!\!
z_{s,k,m}^{\rm cps}(t)
{\rm e}^{-j2\pi r\Delta f t}
{\rm d}t$. Since the desired satellite-UT link is compensated at the UT and the OFDM
parameters are properly designed, the residual Doppler variation within one
OFDM symbol is negligible. Hence,
$\mathbf H_{s,k}(t,q\Delta f)\!=\!
\mathbf H_{s,k}(mT_{\rm sym},q\Delta f)$ within the $m$-th OFDM symbol.
Using the orthogonality among OFDM subcarriers,
\begin{align}
r_{s,k,m,r}
&=
\mathbf b_{s,k}^{H}
\mathbf H_{s,k,m,r}
\mathbf x_{s,m,r}
+
z_{s,k,m,r},\\
\mathbf H_{s,k,m,r}
&=
\mathbf H_{s,k}(mT_{\rm sym},r\Delta f)
=
\mathbf d_{s,k,m,r}\mathbf g_{s,k}^{H},
\end{align}
where we have $\mathbf d_{s,k,m,r}=
\mathbf d_{s,k}(mT_{\rm sym},r\Delta f)=\sum_{\ell=0}^{L_{s,k}-1}
\bar a_{s,k,\ell}
{\rm e}^{j2\pi
\left(
\nu_{s,k,\ell}^{\rm ut}mT_{\rm sym}
-
r\Delta f\tau_{s,k,\ell}^{\rm ut}
\right)}
\mathbf d_{s,k,\ell}$. For simplicity, we omit the OFDM symbol index $m$ and subcarrier
index $r$ in the following: $\mathbf{H}_{s,k} \triangleq \mathbf{d}_{s,k}\mathbf{g}_{s,k}^H$. The transmit signal of the $s$-th satellite is:
\begin{align}
\mathbf x_s
\textstyle=
\sum_{k\in\mathcal K}
\mathbf p_{s,k}x_{s,k}
\in\mathbb C^{M\times 1},
\end{align}
where $\mathbf p_{s,k}\in\mathbb C^{M\times 1}$ denotes the precoding vector
for the $k$-th UT served by the $s$-th satellite, and $x_{s,k}$ is the
corresponding data symbol. Therefore, the received signal from the $s$-th satellite at the $k$-th UT can be written as:
\begin{align}
r_{s,k}
\textstyle=
\mathbf b_{s,k}^{H}
\mathbf H_{s,k}
\sum_{k^{\prime}\in\mathcal K}
\mathbf p_{s,k^{\prime}}x_{s,k^{\prime}}
+
z_{s,k}.
\end{align}
$z_{s,k}$ denotes the aggregate inter-satellite interference-plus-noise term at the $k$-th UT, whose variance is $\mathbb E\{
z_{s,k}z_{s,k}^{*}\}
=
\mathbf b_{s,k}^{H}
\mathbf R_{s,k}
\mathbf b_{s,k}$.
The covariance matrix is:
\begin{align}
\!\!\!\!\mathbf R_{s,k}
\!\textstyle=\!\!\!
\sum_{s^{\prime}\in\mathcal S\!\setminus\!\{\!s\!\}}\!
\!\sum_{k^{\prime}\in\mathcal K}\!
\mathbf g_{s^{\prime},k}^{H}
\mathbf p_{s^{\prime},k^{\prime}}
\mathbf p_{s^{\prime},k^{\prime}}^{H}
\mathbf g_{s^{\prime},k}
\mathbf R_{s^{\prime},k}^{\rm ut}
\!\!+\!\!
\sigma_k^2\mathbf I_N.
\end{align}
Herein, $\mathbf R_{s^{\prime},k}^{\rm ut}$ denotes the
UT-side channel correlation matrix from the interfering satellite $s^{\prime}$ to the $k$-th UT.

\vspace{-4mm}
\subsection{Statistical Characteristics and Angle Error Model}
\vspace{-1mm}
\textcolor{black}{We adopt sCSI rather than instantaneous CSI (iCSI) because the high mobility and non-negligible propagation and feedback delays in LEO systems make timely tracking of fast fading challenging. In contrast, sCSI is governed mainly by slowly varying geometric LoS information and large-scale channel parameters, which can be obtained or predicted from satellite ephemeris, attitude, and UT-reported GNSS information. Owing to the LoS-dominant channel, sCSI-based precoding can retain substantial beamforming gains without tracking instantaneous fading. By exploiting this stable spatial structure, it can also reduce signal leakage toward unintended users and thereby mitigate interference.}

The channel between the $s$-th satellite and the $k$-th UT can be further modeled as a Rician fading channel \cite{Xiang2024,Li2022,wang2025MSMS}:
\begin{align}
\begin{split}
\mathbf{H}_{s,k}\textstyle=\sqrt{\frac{\kappa_{s,k} \beta_{s,k}}{\kappa_{s,k}+1}} \mathbf{d}_{s,k,0}\mathbf{g}_{s,k}^{H}+\sqrt{\frac{\beta_{s,k}}{\kappa_{s,k}+1}} \tilde{\mathbf{d}}_{s,k}\mathbf{g}_{s,k}^{H},
\end{split}
\end{align}
the LoS component satisfies $\|\mathbf{d}_{s,k,0}\|^{2}=\|\mathbf{g}_{s,k}\|^{2}=1$, while the non-LoS (NLoS) component follows a complex Gaussian distribution $\tilde{\mathbf{d}}_{s,k} \sim \mathcal{CN}(\mathbf{0},\boldsymbol{\Sigma}_{s,k})$ with $\operatorname{Tr}(\boldsymbol{\Sigma}_{s,k})=1$. 
The receiver steering vector $\mathbf{d}_{s,k}=\sqrt{\frac{\kappa_{s,k}\beta_{s,k}}{\kappa_{s,k}+1}}\mathbf{d}_{s,k,0} 
+\sqrt{\frac{\beta_{s,k}}{\kappa_{s,k}+1}}\tilde{\mathbf{d}}_{s,k}$,
and $\beta_{s,k}=\textcolor{black}{\mathbb{E}_{\mathbf{d}}}\{\operatorname{Tr}(\mathbf{H}_{s,k}\mathbf{H}_{s,k}^{H})\} 
=\textcolor{black}{\mathbb{E}_{\mathbf{d}}}\{\|\mathbf{d}_{s,k}\|^{2}\}$ \textcolor{black}{denotes the average channel power, which captures the distance-dependent path loss and other large-scale propagation losses, including atmospheric and rain attenuation.} The channel correlation matrices at the UT side and at the satellite side are:
\begin{align}
\!\!\!\!\!\mathbf{R}_{s,k}^{\rm {ut}} \!\!\textstyle=\!\!\textcolor{black}{\mathbb{E}_{\mathbf{d}}}\!\!\left\{\!\mathbf{H}_{s,k} \mathbf{H}_{s,k}^{H}\!\right\}\!\!=\!\!\frac{\kappa_{s,k} \beta_{s,k}}{\kappa_{s,k}\!+\!1} \mathbf{d}_{s,k,0} \mathbf{d}_{s,k, 0}^{H}\!\!+\!\!\frac{\beta_{s,k}}{\kappa_{s,k}\!+\!1} \!\boldsymbol{\Sigma}_{s,k},\label{Rut}
\\
\mathbf{R}_{s,k}^{\rm {sat}}  \textstyle=\textcolor{black}{\mathbb{E}_{\mathbf{d}}}\left\{\mathbf{H}_{s,k}^{H} \mathbf{H}_{s,k}\right\}=\beta_{s,k}\mathbf{g}_{s,k} \mathbf{g}_{s,k}^{H}.
\end{align}
Based on the satellite channel characteristics described above, we summarize the sCSI available at the satellite side: $\{\phi^{\rm sat}_{s,k}, \theta^{\rm sat}_{s,k}, \phi^{\rm ut}_{s,k,0}, \theta^{\rm ut}_{s,k,0}, \boldsymbol{\Sigma}_{s,k}, \,\beta_{s,k}, \kappa_{s,k}\}_{\forall s,\forall k}$. The angle parameters $\{\phi^{\rm sat}_{s,k}, \theta^{\rm sat}_{s,k}, \phi^{\rm ut}_{s,k,0}, \theta^{\rm ut}_{s,k,0}\}_{\forall s,k}$ are primarily obtained from the satellite ephemeris, satellite attitude, and the UT position provided by GNSS, and are used to construct the LoS steering vectors. The average channel power $\beta_{s,k}$, Rician factor $\kappa_{s,k}$, and the normalized NLoS covariance matrix $\boldsymbol{\Sigma}_{s,k}$ can be estimated from the repeated downlink-pilot observations.

The clean and global sCSI commonly assumed in centralized schemes relies on idealized perfect information acquisition or a large number of inter-satellite information exchanges, which are difficult to realize in practical decentralized transmission. Consequently, each satellite typically has access only to partial and noisy observations. 
Since angle information directly determines the precoding structure, we use angle perturbation as a physically meaningful characterization of noisy sCSI, which may arise from angle-estimation errors, satellite attitude jitter, array calibration errors and uncertainty, and feedback delay \cite{Cilden2019, pan2019, Liu2022}. For each link \((s,k)\), the true angle vector and its corresponding error vector are defined as:
\begin{align}
\boldsymbol{\Psi}_{s,k}
&\triangleq
\big[\phi_{s,k}^{\rm sat},\theta_{s,k}^{\rm sat},\phi_{s,k,0}^{\rm ut},\theta_{s,k,0}^{\rm ut}\big]^T,\\
\Delta\boldsymbol{\Psi}_{s,k}
&\triangleq
\big[\Delta\phi_{s,k}^{\rm sat},\Delta\theta_{s,k}^{\rm sat},\Delta\phi_{s,k,0}^{\rm ut},\Delta\theta_{s,k,0}^{\rm ut}\big]^T,
\end{align}
where the four angle error components are assumed to be independent and identically distributed as: $\Delta\phi_{s,k}^{\rm sat},\Delta\theta_{s,k}^{\rm sat},
\Delta\phi_{s,k,0}^{\rm ut},\Delta\theta_{s,k,0}^{\rm ut}
\sim \mathcal{U}[-\varepsilon,\varepsilon]$.
Here, $\varepsilon>0$ represents the angle error bound, i.e., the maximum absolute deviation of each angle component, with a larger $\varepsilon$ indicating less accurate sCSI.
Then, the perturbed angle vector is:
\begin{align}
\bar{\boldsymbol{\Psi}}_{s,k}
\triangleq
\boldsymbol{\Psi}_{s,k}+\Delta\boldsymbol{\Psi}_{s,k}
=
\big[\bar{\phi}_{s,k}^{\rm sat},\bar{\theta}_{s,k}^{\rm sat},\bar{\phi}_{s,k,0}^{\rm ut},\bar{\theta}_{s,k,0}^{\rm ut}\big]^T.
\end{align}
Accordingly, the steering vectors used in student-side statistical modeling are recomputed as:
\begin{align}
\!\!\!\bar{\mathbf g}_{s,k}
&\!=\!
\mathbf a_{M_{\rm x}}\!\Big(\sin \bar{\theta}^{\rm sat}_{s,k}\cos \bar{\phi}^{\rm sat}_{s,k}\Big)
\otimes
\mathbf a_{M_{\rm y}}\!\Big(\cos \bar{\theta}^{\rm sat}_{s,k}\Big),\\
\!\!\!\bar{\mathbf d}_{s,k,0}
&\!=\!
\mathbf a_{N_{\rm x'}}\!\Big(\sin \bar{\theta}^{\rm ut}_{s,k,0}\cos \bar{\phi}^{\rm ut}_{s,k,0}\Big)
\!\otimes\!
\mathbf a_{N_{\rm y'}}\!\Big(\cos \bar{\theta}^{\rm ut}_{s,k,0}\Big).
\end{align}
Then, all sCSI-dependent covariance terms are updated as:
\begin{align}
\bar{\mathbf R}_{s,k}^{\rm ut}
&\textstyle=
\frac{\kappa_{s,k}\beta_{s,k}}{\kappa_{s,k}+1}\,
\bar{\mathbf d}_{s,k,0}\bar{\mathbf d}_{s,k,0}^H
+
\frac{\beta_{s,k}}{\kappa_{s,k}+1}\,\boldsymbol{\Sigma}_{s,k},\\
\bar{\mathbf R}_{s,k}^{\rm sat}
&\textstyle=
\beta_{s,k}\,\bar{\mathbf g}_{s,k}\bar{\mathbf g}_{s,k}^H.
\end{align}

\vspace{-2mm}
\subsection{Closed-Form Structure for Multi-Satellite Transmission}
The DL ergodic rate from the satellite $s$ to the UT $k$ is:
\begin{align}
R_{s,k}&=\mathbb{E}_{\mathbf{H}}\left\{\log_2 \left(1+{{\alpha}_{s,k}^{-1}\mathbf{b}_{s,k}^{H} \mathbf{H}_{s,k} \mathbf{p}_{s,k}\mathbf{p}_{s,k}^H\mathbf{H}_{s,k}^H\mathbf{b}_{s,k}}\right) \right\}\nonumber\\
&=\textcolor{black}{\mathbb{E}_{\mathbf{d}}}\left\{\log_2 \left(1+{\alpha_{s,k}^{-1} |\mathbf{p}_{s,k}^H\mathbf{g}_{s,k}|^2|\mathbf{b}_{s,k}^H\mathbf{d}_{s,k}|^2}\right) \right\},
\end{align}
herein, we have ${\alpha}_{s,k}\!\triangleq\! \sum_{k^{\prime} \in \mathcal{K}\setminus \{k\}}\!|\mathbf{p}_{s,k^{\prime}}^H\mathbf{g}_{s,k}|^2|\mathbf{b}_{s,k}^H\mathbf{d}_{s,k}|^2\!\!+\mathbf b_{s,k}^{H}
\mathbf R_{s,k}
\mathbf b_{s,k}$. 
Assuming the sCSI available at satellites, the WSR maximization problem for multi-satellite transmission can be formulated as \cite{Xiang2024}:
\begin{align}
	\begin{split}
	\max_{\left\{\mathbf{b}_{s,k},\mathbf{p}_{s,k}\right\}_{\forall s, k}}&\quad \textstyle\sum_{s \in \mathcal S} \sum_{k \in \mathcal{K}}  a_{s,k}R_{s,k}, \\
		\text{ s.t. } 
		& \quad \textstyle\sum_{k \in \mathcal{K}} \|\mathbf{p}_{s,k}\|_2^2 \leq P_s^{\rm sat},  \forall s \in  \mathcal S,
	\end{split}
\end{align}
$a_{s,k}$ denotes the user rate weight, and $P_s^{\rm sat}$ is the satellite transmit power budget. To facilitate the optimization, we optimize the WSR maximization problem by solving the weighted minimum mean square error (WMMSE) problem \cite{Shi2011}:
\begin{align}
	&\!\!\!\underset{\left\{{u}_{s,k},{w}_{s,k},\mathbf{p}_{s,k},\mathbf{b}_{s,k}\right\}_{\forall s, k}}{\min} \quad \textstyle\sum_{s \in \mathcal S} \sum_{k \in \mathcal{K}} \left( w_{s,k}e_{s,k}-\log w_{s,k}\right), \nonumber\\
&\quad\quad\text{ s.t. } \quad
		  \textstyle\sum_{k \in \mathcal{K}} \|\mathbf{p}_{s,k}\|_2^2 \leq P_s^{\rm sat},  \forall s \!\in\!  \mathcal S,\!\!\!
\end{align}
where $w_{s,k}$ represents the error weight, and the MSE $e_{s,k}$ is:
\begin{align}
	\begin{split}
e_{s,k}&=\mathbb{E}_{\mathbf{H},{x},{z}}\left\{|u_{s,k}r_{s,k}-x_{s,k}|^2\right\}\\
&=\left|u_{s, k}\right|^{2} \left(\eta_{s, k}+\zeta_{s,k}\right)-u_{s, k}^*\xi_{s, k}^*-u_{s, k}\xi_{s, k}+1,
 \end{split}
\end{align}
where the desired signal level $\xi_{s,k} =  \mathbb{E}_\mathbf{H}\{\mathbf{b}_{s, k}^{H} \mathbf{H}_{s, k} \mathbf{p}_{s, k}\}=\sqrt{\frac{\kappa_{s,k}\beta_{s,k}}{\kappa_{s,k}+1}}\mathbf{b}_{s, k}^{H} \mathbf{d}_{s,k,0}\mathbf{g}_{s,k}^H \mathbf{p}_{s,k}$, and the desired signal power $\zeta_{s,k}=\mathbb{E}_\mathbf{H}\{|\mathbf{b}_{s, k}^{H} \mathbf{H}_{s, k} \mathbf{p}_{s, k}|^2\}=|\mathbf{p}_{s,k}^H\mathbf{g}_{s,k}|^2\mathbf{b}_{s,k}^H\mathbf{R}_{s,k}^{\rm ut}\mathbf{b}_{s,k}$. Moreover, the aggregate interference-plus-noise signal power $\eta_{s, k}=\mathbb{E}_{\mathbf{H},z}\{\sum_{k^{\prime}\in \mathcal{K}\setminus \{k\}}|\mathbf{b}_{s, k}^{H} \mathbf{H}_{s, k} \mathbf{p}_{s, k^{\prime}}|^2+z_{s,k}z_{s,k}^*\}=\sum_{k^{\prime} \in \mathcal{K}\setminus \{k\}}|\mathbf{p}_{s,k^{\prime}}^H\mathbf{g}_{s,k}|^2\mathbf{b}_{s,k}^H\mathbf{R}_{s,k}^{\rm ut}\mathbf{b}_{s,k}+\mathbf{b}_{s,k}^H\mathbf{R}_{s,k}\mathbf{b}_{s,k}$.
The derivation follows the standard WMMSE equivalence and the Karush–Kuhn–Tucker (KKT)-based block-coordinate updates \cite{cao2026deep}, and the closed-form expressions can be summarized as:
\begin{align}
&\mathbf{b}_{s,k}^{\star}
\textstyle=
\sqrt{\frac{\kappa_{s,k}\beta_{s,k}}{\kappa_{s,k}+1}}
\, w_{s,k}u_{s,k}
\Big(
w_{s,k}|u_{s,k}|^2
\sum_{s^{\prime}\in \mathcal{S}}\sum_{k^{\prime}\in \mathcal{K}}\nonumber\\
&
|\mathbf{p}_{s^{\prime},k^{\prime}}^H\mathbf{g}_{s^{\prime},k}|^2
\mathbf{R}_{s^{\prime},k}^{\rm ut}
\!+\!
w_{s,k}|u_{s,k}|^2\sigma_k^2\mathbf{I}_N
\!\Big)^{-1}\!\!\!\!
\mathbf{d}_{s,k,0}\mathbf{g}_{s,k}^H\mathbf{p}_{s,k},
\label{beamforming_cf}
\\
&\mathbf{p}_{s,k}^{\star}
\textstyle=
\sqrt{\frac{\kappa_{s,k}\beta_{s,k}}{\kappa_{s,k}+1}}
\, w_{s,k}u_{s,k}^{*}
\Big(
\sum_{s^{\prime}\in \mathcal{S}}\sum_{k^{\prime}\in \mathcal{K}}
w_{s^{\prime},k^{\prime}}|u_{s^{\prime},k^{\prime}}|^2\nonumber\\
&
\mathbf{b}_{s^{\prime},k^{\prime}}^H\mathbf{R}_{s,k^{\prime}}^{\rm ut}\mathbf{b}_{s^{\prime},k^{\prime}}
\mathbf{g}_{s,k^{\prime}}\mathbf{g}_{s,k^{\prime}}^H
+
\lambda_s\mathbf{I}_M
\Big)^{-1}
\mathbf{g}_{s,k}\mathbf{d}_{s,k,0}^H\mathbf{b}_{s,k},
\label{precoding_cf}
\end{align}
where \(\{\lambda_s\}_{\forall s}\) denote the Lagrange multipliers associated with the satellite transmit-power budget constraints. 

The above updates reveal an explicit mapping from compact low-dimensional variables to the high-dimensional beamforming and precoding vectors. Specifically, once the weights, Lagrange multipliers, and sCSI are available, \(\mathbf{b}_{s,k}\) and \(\mathbf{p}_{s,k}\) can be recovered through the structured matrix-inversion forms in \eqref{beamforming_cf} and \eqref{precoding_cf}. 
Therefore, instead of directly learning the high-dimensional precoders, such a WMMSE-induced structure is exploited for developing a neural recovery framework to predict the compact variables and reconstruct the corresponding precoding solution. 
\section{Multi-Satellite Cooperation: From Teacher NN to Student NNs}\label{Multi-Satellite_Cooperation}
Centralized multi-satellite precoding faces two practical challenges: acquiring global and clean sCSI from all satellites is difficult, while distributing the jointly optimized precoders through ISLs incurs substantial signaling overhead and delay. In contrast, decentralized precoding avoids centralized information collection and precoder dissemination but suffers from degraded performance because each satellite has access only to partial and noisy observations. To bridge this gap, we propose the centralized-training and decentralized-execution KD framework. During offline ground-based training, a centralized teacher NN learns from emulated global and clean sCSI and transfers its cooperative precoding knowledge to lightweight student NNs trained with partial and noisy inputs. The trained student NNs are then uploaded to their corresponding satellites for online onboard inference. During operation, only the student NNs are executed, with each satellite independently inferring its local precoder from locally available sCSI and limited ISL-exchanged information.
\subsection{Knowledge Discovery for Multi-Satellite Transmission}
To reduce the learning burden of precoder recovery by NNs, we reformulate the closed-form precoding solution as follows:
\begin{align}
\!\!\!\!\!\mathbf{p}_{s, k}\!\!=\!\! w_{s, k} u_{s, k}^*\!
\Big(\!
\sum_{k^{\prime}\in \mathcal{K}}
\mathbf{g}_{s,k^{\prime}}\mathbf{g}_{s,k^{\prime}}^H
{\varrho}_{s,k^{\prime}}\!\!+\!\!\lambda_s\mathbf{I}_{M}
\!\Big)^{-1}\!\!\!\!\!\!
\mathbf{g}_{s, k}\bar{\mathbf{d}}_{s, k}^H\mathbf{b}_{s, k}.
\label{re_precoding}
\end{align}
Here, we introduce the low-dimensional variable
\(
{\varrho}_{s,k^{\prime}}\triangleq
\sum_{s^{\prime} \in  \mathcal{S}}w_{s^{\prime},k^{\prime}}|u_{s^{\prime},k^{\prime}}|^2
\mathbf{b}_{s^{\prime},k^{\prime}}^H \mathbf{R}_{s,k^{\prime}}^{\rm ut}\mathbf{b}_{s^{\prime},k^{\prime}},
\forall s,k^{\prime}
\),
and denote
\(
\bar{\mathbf{d}}_{s,k}
\triangleq
\mathbb{E}\{\mathbf{d}_{s,k}\}
=
\sqrt{\frac{\kappa_{s,k}\beta_{s,k}}{\kappa_{s,k}+1}}
\mathbf{d}_{s,k,0}
\).
It follows from \eqref{re_precoding} that the high-dimensional precoding vectors are governed by a set of compact variables, including
\(
\{w_{s,k},u_{s,k}\}_{\forall s,k}
\),
\(
\{\lambda_s\}_{\forall s}
\),
\(
\{\varrho_{s,k^{\prime}}\}_{\forall s,k^{\prime}}
\),
and
\(
\{\mathbf{b}_{s,k}\}_{\forall s,k}
\).
Therefore, instead of directly learning the mapping from channel observations to high-dimensional precoding vectors, we learn the mapping to these compact variables and recover the precoders according to \eqref{re_precoding}. This knowledge-guided reformulation significantly reduces the representation complexity of the NN and provides a more structured and interpretable learning target for multi-satellite transmission.

The accurate prediction of these compact variables depends not only on the NN itself but also on the informativeness and reliability of the input features. An ideal centralized learner can exploit clean global sCSI to learn the coupled variables in \eqref{re_precoding}. 
In practical decentralized systems, however, each satellite is equipped with lightweight NNs due to onboard computation, storage, and real-time inference constraints, and can directly access only partial and noisy sCSI. 
To address this information asymmetry, we propose a teacher-student learning framework. The centralized teacher NN is trained with global and clean observations to discover the compact transmission knowledge embedded in the WMMSE-inspired structure. Each decentralized student NN is designed for lightweight onboard inference and operates with partial and noisy sCSI, together with limited information exchanged through ISLs. Through KD, the informative representations and transmission knowledge learned by the teacher are transferred to the students, thereby bridging the gap between centralized knowledge discovery and decentralized precoder recovery. The detailed designs of the centralized teacher NN and decentralized student NNs are presented in the following subsections.

\vspace{-4mm}
\subsection{Large Centralized Teacher NN for MSMS}\label{teacher_network}
\vspace{-1mm}
\begin{figure}[!t]
		\centering  
\includegraphics[width=0.74\linewidth,trim=0.2cm 0.1cm 0.1cm 0.1cm,clip]{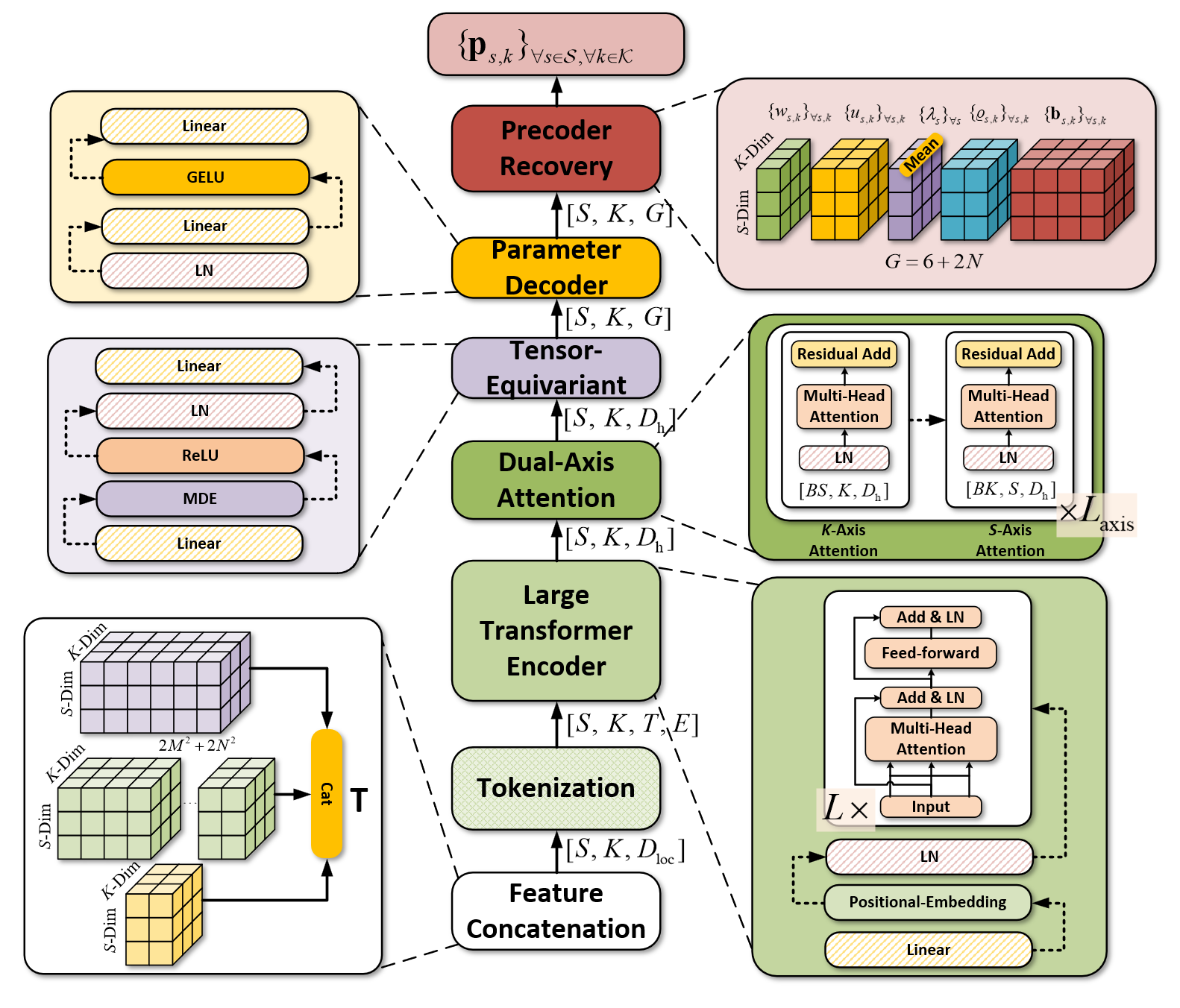}
\caption{Teacher NN framework.}
\label{fig:teacher_network}
\vspace{-4mm}
\end{figure}

According to \eqref{Rut} and \eqref{re_precoding}, we define the input set, which serves as the foundation for the subsequent learning process. In the centralized satellite network architecture, the central satellite combines the local features with the transmitted features from all cooperative satellites, forming the following centralized NN input set $\mathcal{H}^{\rm tea} \triangleq\left\{\mathcal{H}_{s}\right\}_{\forall s \in \mathcal{S}}$:
\begin{align}
\!\!\!\!\mathcal{H}^{\rm tea}
\!\!\triangleq\!\!\left\{ \!\left\{\!\boldsymbol{\Psi}_{s,k},\!\beta_{s,k},\! \kappa_{s,k}, \!\boldsymbol{\Sigma}_{s,k},\!\sigma_{k}^2\!\right\}_{\!\forall k \in \mathcal{K}} \cup  \left\{\!P_s^{\rm sat}\!\right\}\!\right\}_{\!\forall s \in \mathcal S}\!.
\end{align}
 The central teacher satellite feature set $\mathcal{H}^{\rm tea}$ comprises the angle information of the satellite and UTs, the average channel power $\beta_{s,k}$, the Rician factor $\kappa_{s,k}$, the spatial covariance matrix $\boldsymbol{\Sigma}_{s,k}$ of the NLoS component, and the receiver noise power $\sigma_{k}^2$.
Based on \(\mathcal{H}^{\rm tea}\), we construct the centralized teacher NN input tensor $\tT\in \mathbb{R}^{S \times K \times D_{\rm loc}}$:
\begin{align}
\tT_{[s,k,:]} &\!=\! \big[ \phi^{\rm sat}_{s,k},\, \theta^{\rm sat}_{s,k},\, \phi^{\rm ut}_{s,k,0},\, \theta^{\rm ut}_{s,k,0},\beta_{s,k},\kappa_{s,k}, \sigma^2_{s,k}, P_{s,k}^{\rm sat},\nonumber\\ 
&\hspace{1.1em}\operatorname{vec}(\Re(\mathbf{R}_{s,k}^{\rm ut})),\,
\operatorname{vec}(\Im(\mathbf{R}_{s,k}^{\rm ut})),\nonumber\\
&\hspace{1.1em}\operatorname{vec}(\Re(\mathbf{R}_{s,k}^{\rm sat})),\,
\operatorname{vec}(\Im(\mathbf{R}_{s,k}^{\rm sat})) \big],
\label{tensor_T}
\end{align}
where $\sigma_{s,k}^2 \triangleq \sigma_k^2$ and $P_{s,k}^{\rm sat} \triangleq P_{s}^{\rm sat}$.
\subsubsection{\textbf{Large Transformer Encoder (LTE)}}
For each satellite $s\in\mathcal{S}$ and user $k\in\mathcal{K}$, the feature vector $\mathbf{t}_{s,k}\triangleq\tT_{[s,k,:]}\in\mathbb{R}^{D_{\mathrm{loc}}}$ is partitioned into patch tokens of length $E$. With zero-padding of the last patch if necessary, the number of tokens is $T=\left\lceil\frac{D_{\mathrm{loc}}}{E}\right\rceil$,
and the raw patch tensor is
$\tF^{\rm tea}=\mathrm{Token}(\tT) \in\mathbb{R}^{S\times K\times T\times E}$.
The multi-layer Transformer $N_{\mathrm{LTE}}$ serves as the backbone that maps raw satellite-user features into compact patch-conditioned embeddings through stacked self-attention:
\begin{align}
\tX^{\rm tea}=N_{\mathrm{LTE}}^{\mathrm{tea}}(\tF^{\rm tea}) \in \mathbb{R}^{S \times K\times D_{\rm h}}.
\end{align}
Each patch token is linearly projected to hidden dimension $D_{\mathrm{h}}$, summed with a learnable positional embedding that depends on the patch index $\tau$, and applies the layer normalization (LN):
\begin{align}
\mathbf{x}^{(0)}_{s,k,\tau}
\!=\!
\mathrm{LN}\big(\mathrm{Linear}(\mathbf{f}_{s,k,\tau};
\mathbf{W}_{\mathrm{emb}},
\mathbf{b}_{\mathrm{emb}}
)+\!
\mathbf{p}_{\tau}
\big)\!\in\!\mathbb{R}^{D_{\mathrm{h}}},
\end{align}
where $\mathbf{f}_{s,k,\tau}\triangleq \tF^{\mathrm{tea}}_{[s,k,\tau,:]}\in\mathbb{R}^{E},\tau\in\{1,\ldots,T\}$, and $\mathbf{W}_{\mathrm{emb}}\in\mathbb{R}^{E\times D_{\mathrm{h}}}$, $\mathbf{b}_{\mathrm{emb}},\mathbf{p}_{\tau}\in\mathbb{R}^{D_{\mathrm{h}}}$ shared across all $(s,k)$.
We define a multi-dimensional linear layer that applies an affine mapping along the last dimension:
$
\mathrm{Linear}(\tX;\mathbf W,\mathbf b)
\triangleq 
\tX \times \mathbf W + \mathbf{1}_{M_1,\ldots,M_N,1}\odot \mathbf b^{T}$,
where $\tX\in\mathbb{R}^{M_1\times\cdots\times M_N\times D_{\rm I}}$, $\mathbf W\in\mathbb{R}^{D_{\rm I}\times D_{\rm O}}$, and $\mathbf b\in\mathbb{R}^{D_{\rm O}}$.
For each fixed $(s,k)$, the patch embeddings are stacked into a length-$T$ sequence
\begin{align}
\mathbf{X}^{(0)}_{s,k}
=
\left[\mathbf{x}^{(0)}_{s,k,1},\ldots,\mathbf{x}^{(0)}_{s,k,T}\right]^{T}
\in\mathbb{R}^{T\times D_{\mathrm{h}}}.
\end{align}
The resulting sequence is then fed into an \(L\)-layer Transformer encoder, where self-attention is restricted to the patch dimension \(\tau\) for each satellite-user pair:
\begin{align}
\mathbf{X}^{(\ell)}_{s,k}
=
\mathrm{TF}_{\ell}\!\left(\mathbf{X}^{(\ell-1)}_{s,k}\right)\in\mathbb{R}^{T\times D_{\mathrm{h}}}, \quad \forall \ell=1,\ldots,L,
\end{align}
where each layer $\mathrm{TF}_{\ell}(\cdot)$ consists of multi-head attention (MHA) followed by a feed-forward network (FFN).

\textbf{(i) MHA.} Let $\mathbf{Q}=\mathbf{X}^{(\ell-1)}_{s,k}\mathbf{W}^{Q},
\mathbf{K}=\mathbf{X}^{(\ell-1)}_{s,k}\mathbf{W}^{K},
\mathbf{V}=\mathbf{X}^{(\ell-1)}_{s,k}\mathbf{W}^{V}$,
with $\mathbf{W}^{Q},\mathbf{W}^{K},\mathbf{W}^{V}\in\mathbb{R}^{D_{\mathrm{h}}\times D_{\mathrm{h}}}$.
Split into $H$ heads with $d_{\mathrm{k}}=D_{\mathrm{h}}/H$. For head $h$, softmax is taken over the key dimension indexed by patch $\tau$:
\begin{align}
\mathbf{O}_{h}\textstyle=\mathrm{softmax}\!\left({\mathbf{Q}_{h}\mathbf{K}_{h}^{T}}/{\sqrt{d_{\mathrm{k}}}}\right)\mathbf{V}_{h}.
\end{align}
Concatenate heads along the feature dimension,
$\mathbf{O}=[\mathbf{O}_{1},\ldots,\mathbf{O}_{H}]_2\in\mathbb{R}^{T\times D_{\mathrm{h}}}$,
apply an output linear map, dropout, residual from $\mathbf{X}^{(\ell-1)}_{s,k}$, and LN:
\begin{align}
\widetilde{\mathbf{X}}^{(\ell)}_{s,k}
\!=\!
\mathrm{LN}\!\left(
\mathbf{X}^{(\ell-1)}_{s,k}\!+\!\mathrm{Dropout}\!\left(
\mathbf{O}(\mathbf{W}^{O})^T
\right)
\right)\!\in\!\mathbb{R}^{T\times D_{\mathrm{h}}}.
\end{align}

\textbf{(ii) FFN.} A two-layer FFN is applied to every patch row of $\widetilde{\mathbf{X}}^{(\ell)}_{s,k}$ with hidden width $D_{\mathrm{ffn}}=\rho D_{\mathrm{h}}$, ReLU activation \cite{Nair2010_ReLU}, dropout, a residual connection from $\widetilde{\mathbf{X}}^{(\ell)}_{s,k}$, and a final LN:
\begin{align}
\begin{aligned}
\mathbf{X}^{(\ell)}_{s,k}
&=
\mathrm{LN}\Bigg(
\widetilde{\mathbf{X}}^{(\ell)}_{s,k}
+
\mathrm{Dropout}\Bigg(
\mathrm{Linear}\Big(
\mathrm{ReLU}\Big(\\[-0.7em]
&\hspace{2em}
\mathrm{Linear}\big(
\widetilde{\mathbf{X}}^{(\ell)}_{s,k};
\mathbf{W}_{1}^{\rm ffn},\mathbf{b}_{1}^{\rm ffn}\big)
\Big);
\mathbf{W}_{2}^{\rm ffn},\mathbf{b}_{2}^{\rm ffn}
\Big)
\Bigg)
\Bigg),
\end{aligned}
\end{align}
where \(\mathbf{X}^{(\ell)}_{s,k}\in\mathbb{R}^{T\times D_{\mathrm{h}}}\), \(\mathbf{W}_{1}^{\rm ffn}\in\mathbb{R}^{D_{\mathrm{h}}\times D_{\mathrm{ffn}}}\), \(\mathbf{W}_{2}^{\rm ffn}\in\mathbb{R}^{D_{\mathrm{ffn}}\times D_{\mathrm{h}}}\), and \(\mathbf{b}_{1}^{\rm ffn}\in\mathbb{R}^{D_{\mathrm{ffn}}}\), \(\mathbf{b}_{2}^{\rm ffn}\in\mathbb{R}^{D_{\mathrm{h}}}\) are the corresponding biases.
Patch-wise mean pooling yields one vector per $(s,k)$:
\begin{align}
\tX^{\rm tea}_{[s,k,:]}
\textstyle=
\frac{1}{T}\sum_{\tau=1}^{T}\mathbf{x}^{(L)}_{s,k,\tau}\in\mathbb{R}^{D_{\mathrm{h}}},
\end{align}
where $\mathbf{x}^{(L)}_{s,k,\tau} \triangleq  {\mathbf{X}^{(L)}_{s,k}}_{[\tau,:]}$, forming $\tX^{\rm tea}\in\mathbb{R}^{S\times K\times D_{\mathrm{h}}}$ as input to subsequent modules.
\begin{algorithm}[t]
  \caption{Centralized Teacher NN for MSMS}
  \label{net-wmmse-centralized}
  \begin{algorithmic}[1]
    \STATE \textbf{Input:}$\{\boldsymbol{\Psi}_{s,k},\!\beta_{s,k},\!\kappa_{s,k},\!\boldsymbol{\Sigma}_{s,k}\}_{\forall s,k}$,$\{\sigma_{k}^2\}_{\forall k}$,$\{P_s^{\rm sat}\}_{\forall s}$.
    \STATE Construct the centralized teacher NN input tensor $\tT$.
    \STATE \textbf{Training:}
    \STATE Initialize $\mathcal{N}_{\rm tea}^{(0)}=\{N_{\rm LTE}^{\rm tea},N_{\rm DAA}^{\rm tea},N_{\rm TEN}^{\rm tea},N_{\rm PD}^{\rm tea}\}^{(0)}$, $n\!=\!0$.
    \STATE \textbf{for} $n < N_{\rm ep}$ \textbf{do}
      \STATE \quad \textbf{forward:}
        \STATE \quad\quad Tokenize the input: $\tF^{\rm tea}=\mathrm{Token}(\tT) $.
      \STATE \quad\quad Apply the LTE network: $\tX^{\rm tea} = N_{\rm LTE}^{\rm tea}\big(\tF^{\rm tea}\big)$.
      \STATE \quad\quad Apply the DAA network: $\tH^{\rm tea} = N_{\rm DAA}^{\rm tea}\big(\tX^{\rm tea}\big)$.
    \STATE \quad\quad Apply the TEN network: $\tE^{\rm tea} = N_{\rm TEN}^{\rm tea}\big(\tH^{\rm tea}\big)$.
      \STATE \quad\quad Execute the PD module: $\tC^{\rm tea}=N_{\mathrm{PD}}^{\rm tea}(\tE^{\rm tea})$.
      \STATE \quad\quad $\{\mathbf{b}_{s,k}^{\rm tea},\mathbf{p}_{s,k}^{\rm tea}\}_{\forall s, k}=N_{\mathrm{PR}}^{\rm tea}(\tC^{\rm tea},\{\mathbf{d}_{s,k,0},\mathbf{g}_{s,k}\}_{\forall s, k })$.
      \STATE \quad \textbf{backward:}
      \STATE \quad\quad Update $\mathcal{N}_{\rm tea}^{(n)}$ via gradient descent on $\mathcal{L}_{\rm obj}^{\rm wsr}$. 
      \STATE \quad \quad Update $n=n+1$.
    \STATE \textbf{end for}
   \STATE Store the learned variables of $\mathcal{N}_{\rm tea}$.
    \STATE \textbf{Inference:}  Use $\mathcal{N}_{\rm tea}$ to obtain $\{\mathbf{p}_{s,k}^{\rm tea},\mathbf{b}_{s,k}^{\rm tea}\}_{\forall s,\forall k}$.
  \end{algorithmic}
\end{algorithm}
\subsubsection{\textbf{Dual-Axis Attention (DAA)}}
The DAA network stacks \(L_{\rm axis}\) identical stages. Each stage executes two separable self-attention operations on the satellite–UT lattice: first, along the UT index \(k\) within each satellite \(s\), so that UTs served by the same satellite attend over one another, then along the satellite index \(s\) while holding \(k\) fixed, aggregating embeddings of the same UT across satellites. The UT-axis sweep encodes intra-satellite multi-UT coupling; the satellite-axis sweep fuses inter-satellite information perceived by individual UTs:
\begin{align}
\tH^{\rm tea}=N_{\mathrm{DAA}}^{\rm tea}(\tX^{\rm tea}) \in \mathbb{R}^{S \times K\times D_{\rm h}},
\end{align}
\(N_\mathrm{DAA}^{\rm tea}(\cdot)\) denote \(L_{\rm axis}\) compositions of two-axis updates.

\textbf{(i) User-axis.}
For each satellite $s\in \mathcal{S}$, form a token matrix whose $k$-th row is the embedding of user $k$ on satellite $s$,
\begin{align}
\mathbf{A}^{(\ell)}_{s}
=
\left[\tH^{(\ell-1)}_{[s,1,:]}
,\ldots,
\tH^{(\ell-1)}_{[s,K,:]}\right]^T\in\mathbb{R}^{K\times D_{\mathrm{h}}},
\end{align}
where $\tH^{(0)}=\tX^{\mathrm{tea}}\in\mathbb{R}^{S\times K\times D_{\mathrm{h}}}$. Update $\mathbf{A}^{(\ell)}_{s}$ with LN, MHA along the row index $k$, dropout, and a residual connection,
\begin{align}
\mathbf{A}^{(\ell)}_{s}
\leftarrow
\mathbf{A}^{(\ell)}_{s}
+\mathrm{Dropout}\!\left(
\mathrm{MHA}^{(\ell)}_{\mathrm{user}}\!\big(\mathrm{LN}(\mathbf{A}^{(\ell)}_{s})\big)
\right).
\end{align}
Let $\acute{\tH}^{(\ell)}\in\mathbb{R}^{S\times K\times D_{\mathrm{h}}}$ pack these slices, $\acute{\tH}^{(\ell)}_{[s,k,:]}={\mathbf{A}^{(\ell)}_{s}}_{[k,:]}$.

\textbf{(ii) Satellite-axis.}
For each user $k\in \mathcal{K}$, form a token matrix whose $s$-th row is that user's embedding on satellite $s$ after user-axis mixing,
\begin{align}
\mathbf{B}^{(\ell)}_{k}
=\left[
\acute{\tH}^{(\ell)}_{[1,k,:]},\ldots,\acute{\tH}^{(\ell)}_{[S,k,:]}\right]^T\in\mathbb{R}^{S\times D_{\mathrm{h}}},
\end{align}
and update with the same pattern along the row index $s$,
\begin{align}
\mathbf{B}^{(\ell)}_{k}
\leftarrow
\mathbf{B}^{(\ell)}_{k}
+\mathrm{Dropout}\!\left(
\mathrm{MHA}^{(\ell)}_{\mathrm{sat}}\!\big(\mathrm{LN}(\mathbf{B}^{(\ell)}_{k})\big)
\right).
\end{align}
Define the stage output $\tH^{(\ell)}\in\mathbb{R}^{S\times K\times D_{\mathrm{h}}}$ by $\tH^{(\ell)}_{[s,k,:]}={\mathbf{B}^{(\ell)}_{k}}_{[s,:]}$.
After $L_{\rm axis}$ stages,
$\tH^{\rm tea}\triangleq\tH^{(L_{\rm axis})}$ outputs to subsequent modules.

\subsubsection{\textbf{Tensor-Equivariant Neural (TEN) Network}}
We employ a low-complexity TEN network to transform the input tensor into a latent representation while preserving permutation equivariance with respect to both the satellite and UT dimensions. Specifically, TEN captures both original and multidimensional global features by aggregating mean features across different subsets of the equivariant dimensions and combining them with the original features via learnable linear transformations \cite{wang2024}:
\begin{align}
\tE^{\rm tea}=N_{\mathrm{TEN}}^{\rm tea}(\tH^{\rm tea}) \in \mathbb{R}^{S \times K\times G}.
\end{align}
The TEN network $N_{\mathrm{TEN}}^{\rm tea}$ consists of an input linear layer followed by \(L_{\rm ten}\) stacked blocks, each comprising a multi-dimensional equivariant (MDE) module, a ReLU, an LN, and an output linear layer:
\begin{subequations}
\begin{align}
\tE^{(0)}
&=\mathrm{Linear}\!\left(\tH^{\rm tea};\mathbf W_{1}^{\rm ten},\mathbf b_{1}^{\rm ten}\right)\in\mathbb{R}^{S\times K\times D_{\rm ten}},\\
\tE^{(\ell)}
&=\!\operatorname{LN}\!\Big(\!\operatorname{ReLU}\!\big(\operatorname{MDE}_{\ell}(\tE^{(\ell-1)})\big)\!\Big),\ell\!=\!1,\ldots,L_{\rm ten},\\
\tE^{\rm tea}
&=\mathrm{Linear}\!\left(\tE^{(L_{\rm ten})};\mathbf W_{2}^{\rm ten},\mathbf b_{2}^{\rm ten}\right)\in\mathbb{R}^{S\times K\times G},
\end{align}
\end{subequations}
where $D_{\rm ten}$ denotes the hidden dimension in TEN network and $G=6+2N$, \(\mathbf{W}_1^{\rm ten}\in\mathbb{R}^{D_{\rm h}\times D_{\rm ten}}\), \(\mathbf{W}_2^{\rm ten}\in\mathbb{R}^{D_{\rm ten}\times G}\), \(\mathbf{b}_1^{\rm ten}\in\mathbb{R}^{D_{\rm ten}},\mathbf{b}_{2}^{\rm ten}\in\mathbb{R}^{G}\).
In each layer, \(\operatorname{MDE}_{\ell}(\cdot)\) is implemented as a linear combination of mean-and-repeat patterns defined on subsets of the two equivariant dimensions \(\{S, K\}\) \cite{wang2024}. 
\subsubsection{\textbf{Parameter Decoder (PD) and Precoder Recovery (PR)}}
The PD module aims to decode the precoder-related parameters from the teacher-side representation:
\begin{align}
\tC^{\rm tea}=N_{\mathrm{PD}}^{\rm tea}\big(\tE^{\rm tea}\big).
\end{align}
Specifically, we employ a lightweight MLP consisting of one LN layer, two linear transformations, and a GELU activation. The decoded parameter tensor $\tC^{\rm tea}$ is obtained as: 
\begin{subequations}
\begin{align}
\tilde{\tC}=\mathrm{Linear}(\mathrm{LN}(\tE^{\rm tea});\mathbf{W}_1^{\rm pd},\mathbf{b}_1^{\rm pd}),\\
\tC^{\rm tea}=\mathrm{Linear}(\mathrm{GELU}(\tilde{\tC});\mathbf{W}_2^{\rm pd},\mathbf{b}_2^{\rm pd}),
\end{align}
\end{subequations}
with $\mathbf{W}_{1}^{\rm pd},\mathbf{W}_{2}^{\rm pd}\in\mathbb{R}^{G\times G}$ and $\mathbf{b}^{\rm pd}_{1},\mathbf{b}^{\rm pd}_{2}\in\mathbb{R}^{G}$. Then, the PD module recovers the receive and precoding vectors: 
\begin{align}
\{\mathbf{b}_{s,k}^{\rm tea},\mathbf{p}_{s,k}^{\rm tea}\}_{\forall s,\forall k}=N_{\mathrm{PR}}^{\rm tea}\big(\tC^{\rm tea},\{\mathbf{d}_{s,k,0},\mathbf{g}_{s,k}\}_{\forall s,\forall k }\big),
\end{align}
writing $\tC^{\rm tea}_{[s,k,:]}=\mathbf{c}_{s,k}=[c_1,\ldots,c_G]^{T}$,
\begin{align}
\left\{\begin{matrix}
u_{s,k}^{\rm tea}  = c_1 + j c_2, \\
w_{s,k}^{\rm tea}  = \mathrm{softplus}(c_3)+\epsilon,\\
\tilde{\lambda}_{s,k}^{\rm tea}  = c_4,\,\,
\lambda_s^{\rm tea} = \mathrm{softplus}\!\left(
\frac{1}{K}\sum_{k = 1}^{K}\tilde{\lambda}_{s,k}
\right), \\
\varrho_{s,k}^{\rm tea}  = c_5 + j c_6, \\
\mathbf{b}_{s,k}^{\rm tea}  = \mathrm{norm}\!\left(
{\mathbf{c}_{s,k}}_{[7:6+N]}
+j{\mathbf{c}_{s,k}}_{[7+N:6+2N]}
\right). \\
\end{matrix}\right.
\end{align}
Herein, $\mathrm{norm}(\cdot)$ is $\ell_2$ normalization on $\mathbb{C}^{N}$, and $\epsilon>0$ is small.
Given $\{u_{s,k}^{\rm tea},w_{s,k}^{\rm tea},\lambda_{s}^{\rm tea},\varrho_{s,k}^{\rm tea},\mathbf{b}_{s,k}^{\rm tea}\}_{\forall s, \forall k}$ and  $\{\mathbf{d}_{s,k,0},\mathbf{g}_{s,k}\}_{\forall s,\forall k }$, we can recover the closed-form precoders $\{\mathbf{p}_{s,k}^{\rm tea}\}_{\forall s,\forall k}$ as in \eqref{re_precoding}. Note that $N_{\rm PR}^{\rm tea}$ is a fixed computation module and is not updated during training. 
Moreover, the teacher NN $\mathcal{N}_{\rm tea}=\{N_{\rm LTE}^{\rm tea},N_{\rm DAA}^{\rm tea},N_{\rm TEN}^{\rm tea},N_{\rm PD}^{\rm tea}\}$ is frozen during following distillation. 
The entire teacher NN training and inference process is summarized in \algref{net-wmmse-centralized}.

\begin{figure}[!t]
		\centering  
\includegraphics[width=0.74\linewidth,trim=0.1cm 0.1cm 0.1cm 0.1cm,clip]{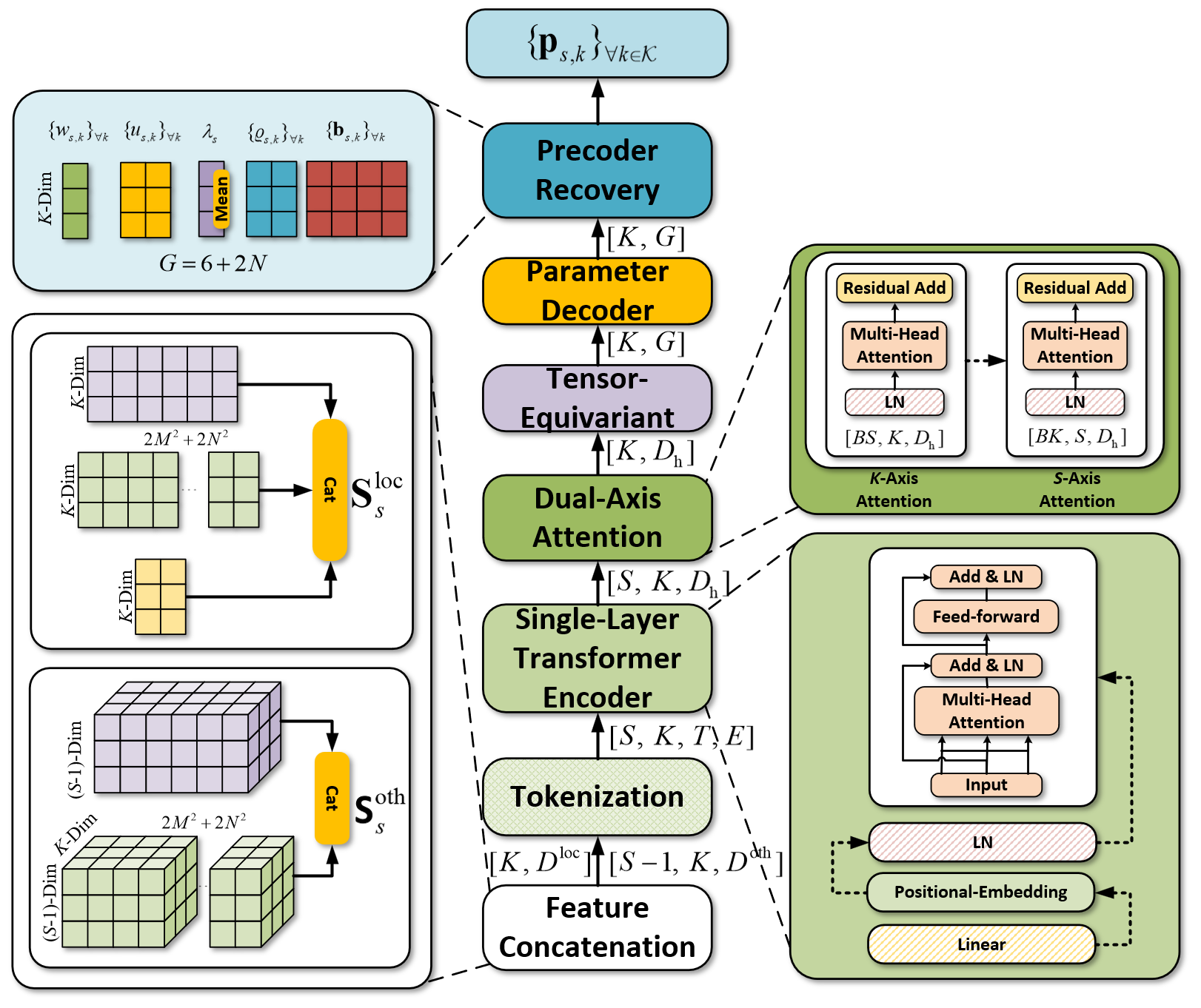}
\caption{Student NN framework.}
\label{fig:student_network}
\vspace{-5mm}
\end{figure}
\subsection{Lightweight Decentralized Student NNs for MSMS}\label{student_network}
The decentralized student is a weight-shared NN instantiated for each satellite $s\in\mathcal{S}$. It maps local features, together with low-rate information exchanged among peer satellites, to per-satellite precoders without centralized aggregation \cite{cao2026deep}. \textcolor{black}{Assuming that only satellite state information (SSI), including the satellite position, attitude, and transmit power, is periodically exchanged among decentralized student satellites, the exchanged information, together with the UT positions, is used to obtain the corresponding angles $\{\bar{\boldsymbol{\Psi}}_{s^{\prime},k}\}_{\forall s \in \mathcal{S}\setminus\{s\},\forall k \in \mathcal{K}}$ for the other satellites. The student feature set is then constructed from the local- and other-satellite features:}
\begin{align}
\mathcal{H}_{s}^{\rm stu} \triangleq &
\left\{\bar{\boldsymbol{\Psi}}_{s,k},\beta_{s,k}, \kappa_{s,k}, \boldsymbol{\Sigma}_{s,k},\sigma_{k}^2,P_s^{\rm sat}\right\}_{\forall k \in \mathcal{K}}\\ &\cup  \{\bar{\boldsymbol{\Psi}}_{s^{\prime},k},\sigma_{k}^2,P_{s^{\prime}}^{\rm sat}\}_{\forall s^{\prime} \in \mathcal S \setminus \{s\}, \forall k \in \mathcal{K}},
\end{align}
Based on $\mathcal{H}_{s}^{\rm stu}$, we construct the local-satellite input second-order tensors
\(\mathbf{S}_s^{\rm loc} \in \mathbb{R}^{K \times D_{\rm loc}}\) and the other-satellite input third-order tensors
\(\tS^{\rm oth}_s \in \mathbb{R}^{(S-1) \times K \times D_{\rm oth}}\) as follows:
\begin{align}
{\mathbf{S}_s^{\rm loc}}_{[k,:]} &\!=\! \big[ \bar{\phi}^{\rm sat}_{s,k},\bar{\theta}^{\rm sat}_{s,k},\bar{\phi}^{\rm ut}_{s,k,0},\bar{\theta}^{\rm ut}_{s,k,0},\beta_{s,k},\kappa_{s,k}, \sigma^2_{s,k}, P_{s,k}^{\rm sat},\nonumber\\ 
&\hspace{1.1em}\operatorname{vec}(\Re(\bar{\mathbf{R}}_{s,k}^{\rm ut})),\
\operatorname{vec}(\Im(\bar{\mathbf{R}}_{s,k}^{\rm ut})),\nonumber\\
&\hspace{1.1em}\operatorname{vec}(\Re(\bar{\mathbf{R}}_{s,k}^{\rm sat})),\,
\operatorname{vec}(\Im(\bar{\mathbf{R}}_{s,k}^{\rm sat})) \big],
\end{align}
\begin{align}
{\tS_s^{\rm oth}}_{[s^{\prime},k,:]} &\!=\! \big[ \bar{\phi}^{\rm sat}_{s^{\prime},k},\bar{\theta}^{\rm sat}_{s^{\prime},k},\bar{\phi}^{\rm ut}_{s^{\prime},k,0},\bar{\theta}^{\rm ut}_{s^{\prime},k,0},\sigma^2_{s^{\prime},k}, P_{s^{\prime},k}^{\rm sat},\nonumber\\ 
&\hspace{1.1em}\operatorname{vec}(\Re(\bar{\mathbf{d}}_{s^{\prime},k,0}\bar{\mathbf{d}}_{s^{\prime},k,0}^H)),
\operatorname{vec}(\Im(\bar{\mathbf{d}}_{s^{\prime},k,0}\bar{\mathbf{d}}_{s^{\prime},k,0}^H)),\nonumber\\
&\!\!\!\!\!\!\!\!\!\!\!\!\!\!\!\!\!\!\!\!\!\!\!\!\operatorname{vec}(\Re(\bar{\mathbf{g}}_{s^{\prime},k}\bar{\mathbf{g}}_{s^{\prime},k}^H)), \operatorname{vec}(\Im(\bar{\mathbf{g}}_{s^{\prime},k}\bar{\mathbf{g}}_{s^{\prime},k}^H)) \big], s^{\prime} \!\in\! \mathcal{S}\! \setminus\! \{\!s\!\},
\end{align}

Fixing a target satellite $s$, we collect the local-satellite feature tensor $\mathbf{S}_s^{\mathrm{loc}}\in\mathbb{R}^{K\times D_{\mathrm{loc}}}$ and other-satellite feature tensor $\tS_s^{\mathrm{oth}}\in\mathbb{R}^{(S-1)\times K\times D_{\mathrm{oth}}}$ for $s^{\prime}\in\mathcal{S}\setminus \{s\}$. Due to the limited observation of the decentralized student satellite, $D_{\mathrm{oth}}= D_{\mathrm{loc}}-2$, we align the feature dimension by zero-padding so that all features are put into $\mathbb{R}^{S\times K\times D_{\mathrm{loc}}}$. We then form a virtual stack along an auxiliary satellite index,
\begin{align}
\tS_{s} \triangleq
\left[\mathbf{S}_s^{\rm loc},\left[\tS_s^{\rm oth},\mathbf{0}_{(S-1)\times K\times 2}
\right]_3\right]_1
\in\mathbb{R}^{ S\times K\times D_{\mathrm{loc}}}.
\label{eq:virtual_stack}
\end{align}
Compared with the teacher-NN input, the student-NN input excludes the average channel power $\beta_{s^{\prime},k}$ and the Rician factor $\kappa_{s^{\prime},k}$ of the other satellites, and all angle information is corrupted by estimation errors, satellite attitude jitter, array calibration errors and uncertainty, and feedback delays \cite{Cilden2019, pan2019, Liu2022}. Moreover, we use the outer products of the LoS steering vectors and their Hermitian transposes, i.e., $\bar{\mathbf{g}}_{s^{\prime},k}\bar{\mathbf{g}}_{s^{\prime},k}^{H}$ and $\bar{\mathbf{d}}_{s^{\prime},k,0}\bar{\mathbf{d}}_{s^{\prime},k,0}^{H}$, as substitutes for the missing correlation matrices.
The teacher in \secref{teacher_network} assumes a global-capable node that acquires the full tensor $\tT\!\in\!\mathbb{R}^{S\times K\times D_{\rm loc}}$ and runs a deep LTE backbone ($L$ Transformer layers).
Each decentralized student satellite, however, must perform on-board precoding under latency and signaling constraints. Relative to the teacher, the LTE is reduced to a single-layer Transformer encoder (STE) for patch tokens, rather than $L$ Transformer layers. The whole student NN for the $s$-th satellite is:
\begin{subequations}
\begin{align}
\tF^{\rm stu}_s &= \mathrm{Token}\big(\tS_s\big)\in\mathbb{R}^{ S\times K\times T \times E},\\
\tX^{\rm stu}_s &= N_{\rm STE}^{\rm stu}\big(\tF^{\rm stu}_s\big) \in\mathbb{R}^{ S\times K\times D_{\mathrm{h}}},\\
\tH^{\rm stu}_s&=N_{\rm DAA}^{\rm stu}\big(\tX^{\rm stu}_s\big)\in\mathbb{R}^{ S\times K\times D_{\mathrm{h}}},\\
{\bar{\mathbf{H}}^{\rm stu}_{s}}&=\frac{1}{S}\sum_{s^{\prime}=1}^{S}\tH^{\rm stu}_{s[s^{\prime},:,:]}
\in\mathbb{R}^{K\times D_{\mathrm{h}}},\\
\mathbf{E}^{\rm stu}_s &=N_{\rm TEN}^{\rm stu}\big(\bar{\mathbf{H}}^{\rm stu}_{s}\big)\in\mathbb{R}^{ K\times G},\\
\mathbf{C}^{\rm stu}_s&=N_{\rm PD}^{\rm stu}\big(\mathbf{E}^{\rm stu}_s\big),\\
\{\mathbf{b}_{s,k}^{\rm stu},\mathbf{p}_{s,k}^{\rm stu}\}_{\forall k}&=N_{\rm PR}^{\rm stu}\big(\mathbf{C}^{\rm stu}_s,\{\bar{\mathbf{d}}_{s,k,0},\bar{\mathbf{g}}_{s,k}\}_{\forall k }\big).
\end{align}
\end{subequations}
 Herein, the student NNs we deployed are weight-shared across different satellites. 
 The student NN training and inference process is summarized in \algref{net-wmmse-decentralized}.
\begin{algorithm}[t]
  \caption{Decentralized Student NNs for MSMS}
  \label{net-wmmse-decentralized}
  \begin{algorithmic}[1]
    \STATE \textbf{Input:} $\left\{\{\bar{\boldsymbol{\Psi}}_{s,k},\beta_{s,k}, \kappa_{s,k}, \boldsymbol{\Sigma}_{s,k},\sigma_{k}^2,P_s^{\rm sat}\}_{\forall k \in \mathcal{K}} \right.$\\
    \quad\quad \quad \quad $\left. \cup  \{\bar{\boldsymbol{\Psi}}_{s^{\prime},k},\sigma_{k}^2,P_{s^{\prime}}^{\rm sat}\}_{\forall s^{\prime} \in \mathcal S \setminus \{s\}, \forall k \in \mathcal{K}}\right\}_{\forall s}$.
    \STATE Construct $\{\mathbf{S}_s^{\rm loc},\tS^{\rm oth}_s\}_{\forall s}$ and stack into $\{\tS_s\}_{\forall s}$.
    \STATE \textbf{Training:}
    \STATE $\mathcal{N}_{\rm stu}^{(0)}=\{N_{\rm STE}^{\rm stu}, N_{\rm DAA}^{\rm stu},N_{\rm TEN}^{\rm stu}, N_{\rm PD}^{\rm stu}\}^{(0)}$ and $n\!=\!0$.
    \STATE \textbf{for} $n < N_{\rm ep}$ \textbf{do}
      \STATE \quad \textbf{forward:}
      \STATE \quad\quad  \textbf{for} each satellite $s \in \mathcal S$ \textbf{do}
      \STATE \quad\quad\quad Tokenize the input: $\tF^{\rm stu}_s = \mathrm{Token}\big(\tS_s\big)$.\\
      \STATE \quad\quad\quad Apply the STE network: $\tX^{\rm stu}_s = N_{\rm STE}^{\rm stu}\big(\tF^{\rm stu}_s\big)$.
    \STATE \quad\quad\quad Apply the DAA network: $\tH^{\rm stu}_s = N_{\rm DAA}^{\rm stu}\big(\tX^{\rm stu}_s\big)$.
     \STATE \quad\quad\quad Calculate: ${\bar{\mathbf{H}}^{\rm stu}_{s}}=\frac{1}{S}\sum_{s^{\prime}=1}^{S}\tH^{\rm stu}_{s[s^{\prime},:,:]}$.
     \STATE \quad\quad\quad Apply the TEN network: $\mathbf{E}^{\rm stu}_s = N_{\rm TEN}^{\rm stu}\big(\bar{\mathbf{H}}^{\rm stu}_s\big)$.
     \STATE \quad\quad\quad Execute the PD module: $\mathbf{C}_s^{\rm stu}= N_{\rm PD}^{\rm stu}\big(\mathbf{E}_s^{\rm stu}\big)$.
     \STATE \quad\quad\quad $\{\mathbf{b}_{s,k}^{\rm stu},\mathbf{p}_{s,k}^{\rm stu}\}_{\forall k}= N_{\rm PR}^{\rm stu}\big(\mathbf{C}_s^{\rm stu},\{\bar{\mathbf{d}}_{s,k,0},\bar{\mathbf{g}}_{s,k}\}_{\forall k}\big)$.
      \STATE \quad\quad \textbf{end for}
      \STATE \quad \textbf{backward:}
      \STATE \quad\quad Update $\mathcal{N}_{\rm stu}^{(n)}$ via gradient descent on $\mathcal{L}_{\rm obj}^{\rm wsr}$. 
     \STATE \quad \quad Update $n=n+1$.
    \STATE \textbf{end for}
   \STATE Store the learned variables of $\mathcal{N}_{\rm stu}$.
    \STATE \textbf{Inference:}  For $\forall s$, use the partial noisy input $\tS_s$
    and the trained $\mathcal{N}_{\rm stu}$ to obtain
    $\{\mathbf p_{s,k}^{\rm stu},\mathbf b_{s,k}^{\rm stu}\}_{\forall k}$ in parallel.
  \end{algorithmic}
\end{algorithm}

\section{Knowledge Distillation Framework Design}
\label{Knowledge_Distillation_Design}

\subsection{Differences between Teacher and Student}
\begin{figure}[!t]
    \centering  \includegraphics[width=0.82\linewidth,trim=0.1cm 0.1cm 0.1cm 0.1cm,clip]{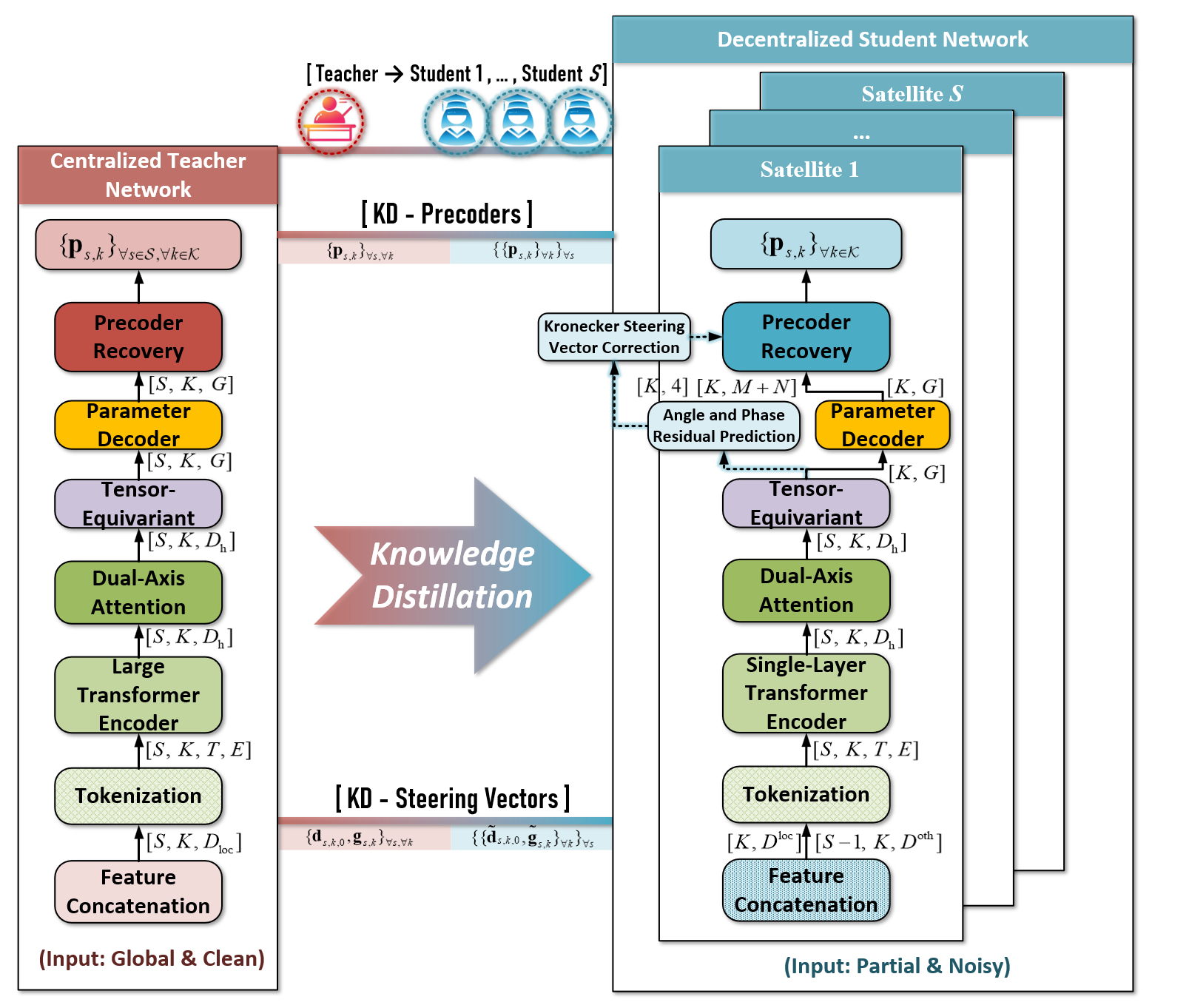}
    \caption{Hybrid knowledge distillation design.}
\label{fig:kd_design_framework}
\vspace{-5mm}
\end{figure}

As illustrated in \figref{fig:kd_design_framework}, the teacher serves as a
privileged knowledge source with access to more complete and reliable observations
and a broader precoding decision scope than each student. Specifically, the teacher
takes the global-clean observation $\tT$ in \eqref{tensor_T} as input, which
contains global features across all satellites and UTs together with accurate angle observations. In contrast, the $s$-th student only has access to
the partial and angle-noisy observation $\tS_s$ in
\eqref{eq:virtual_stack}. Their respective observation domains are characterized as:
\begin{align}
\tT \sim \mathcal{D}_{\rm tea}^{\rm gc},\quad
\tS_s \sim \mathcal{D}_{\rm stu}^{\rm pn},\quad
\mathcal{D}_{\rm tea}^{\rm gc}
\neq
\mathcal{D}_{\rm stu}^{\rm pn},
\end{align}
where $\mathcal{D}_{\rm tea}^{\rm gc}$ denotes the global-clean teacher-view
domain, whereas $\mathcal{D}_{\rm stu}^{\rm pn}$ denotes the partial-noisy
student-view domain.

The teacher-student asymmetry arises from three aspects. First, in terms of
\emph{observation scope}, the teacher observes system-wide information across all
satellites and UTs, whereas each student only observes satellite-specific partial
information. Second, in terms of \emph{observation quality}, the teacher receives
accurate angle observations, whereas the student operates with angle-noisy
observations. Third, in terms of \emph{precoding decision scope}, the centralized teacher learns a system-level mapping from global multi-satellite observations to
jointly coordinated precoders for all satellites, whereas each decentralized student learns a node-level mapping from its partial observation to the precoder of the corresponding satellite. These mappings can be expressed as:
\begin{align}
f_{\rm tea}:\ \tT
&\longmapsto
\left\{
\mathbf{p}^{\rm tea}_{s,k}
\right\}_{\forall s,k},
\\
f_{\rm stu}:\ \tS_s
&\longmapsto
\left\{\mathbf{p}^{\rm stu}_{s,k}\right\}_{\forall k},
\, \forall s\in\mathcal{S},
\end{align}
where $\{
\mathbf{p}^{\rm tea}_{s,k}
\}_{\forall s,k}$ denotes the jointly generated multi-satellite
precoders by the teacher, while $\{\mathbf{p}^{\rm stu}_{s,k}\}_{\forall k}$ denotes the local precoder generated
independently by the $s$-th student satellite.

Accordingly, the proposed framework performs knowledge transfer not only from the
global-clean observation domain to the partial-noisy observation domain, but also
from a centralized system-level precoding policy to decentralized satellite-level
precoding policies. This design enables each student to inherit the teacher's
knowledge of global interference management and multi-satellite coordination while retaining the low signaling overhead and decentralized inference capability
required in practical systems.

\vspace{-4mm}
\subsection{Angle and Phase Correction}
\vspace{-1mm}
To mitigate performance degradation caused by angle noise, we introduce two modules, termed angle and phase residual prediction (APRP) and Kronecker steering-vector correction (KSVC), following the TEN network, as shown in \figref{fig:kd_design_framework}. 
\subsubsection{APRP}
Let $\bar{\boldsymbol{\Psi}}_{s,k}$ denote the noisy angle observation associated with 
the satellite and UT, given by:
\begin{align}
\bar{\boldsymbol{\Psi}}_{s,k}
=
\big[
\bar{\phi}^{\mathrm{sat}}_{s,k},
\bar{\theta}^{\mathrm{sat}}_{s,k},
\bar{\phi}^{\mathrm{ut}}_{s,k,0},
\bar{\theta}^{\mathrm{ut}}_{s,k,0}
\big]^T .
\end{align}
An angle-error prediction head outputs a four-dimensional residual vector
$\mathbf{r}_{s,k}=[r_{s,k}^1,r_{s,k}^2,r_{s,k}^3,r_{s,k}^4]^{T}$, from which the bounded angle corrections are obtained as follows:
\begin{align}
\Delta\hat{\phi}^{\mathrm{sat}}_{s,k} 
&= \alpha_{\phi}\tanh(r_{s,k}^1),
&
\Delta\hat{\theta}^{\mathrm{sat}}_{s,k} 
&= \alpha_{\theta}\tanh(r_{s,k}^2), \nonumber\\
\Delta\hat{\phi}^{\mathrm{ut}}_{s,k,0} 
&= \alpha_{\phi}\tanh(r_{s,k}^3),
&
\Delta\hat{\theta}^{\mathrm{ut}}_{s,k,0} 
&= \alpha_{\theta}\tanh(r_{s,k}^4),
\end{align}
where $\alpha_{\theta}$ and $\alpha_{\phi}$ denote the maximum correction magnitudes 
for the elevation and azimuth angles, respectively.
The satellite-side and UT-side phase-residual vectors are:
\begin{align}
\Delta{\boldsymbol{\psi}}^{g}_{s,k} 
= \psi_{g}\tanh({\mathbf r}_{s,k}^{g}) \in \mathbb{R}^{M \times 1}, \\
\Delta{\boldsymbol{\psi}}^{ d}_{s,k} 
=\psi_{d}\tanh({\mathbf r}_{s,k}^{d}) \in\mathbb{R}^{N \times 1},
\end{align}
$\psi_{g}$ and $\psi_{d}$ are the maximum correction magnitudes 
for the satellite-side and UT-side steering-vector phases, respectively.
\subsubsection{KSVC}
Accordingly, the predicted angle residual gives the corrected angle vector:
$\hat{\boldsymbol{\Psi}}_{s,k}
\triangleq
\bar{\boldsymbol{\Psi}}_{s,k}
-
\Delta\hat{\boldsymbol{\Psi}}_{s,k} =
\big[
\hat{\phi}_{s,k}^{\rm sat},
\hat{\theta}_{s,k}^{\rm sat},
\hat{\phi}_{s,k,0}^{\rm ut},
\hat{\theta}_{s,k,0}^{\rm ut}
\big]^T 
\!\!=\!\!
\big[
\bar{\phi}^{\mathrm{sat}}_{s,k}\!\!-\!\!\Delta\hat{\phi}^{\mathrm{sat}}_{s,k},
\bar{\theta}^{\mathrm{sat}}_{s,k}\!\!-\!\!\Delta\hat{\theta}^{\mathrm{sat}}_{s,k},
\bar{\phi}^{\mathrm{ut}}_{s,k,0}\!\!-\!\!\Delta\hat{\phi}^{\mathrm{ut}}_{s,k,0},
\bar{\theta}^{\mathrm{ut}}_{s,k,0}\!\!-\!\!\Delta\hat{\theta}^{\mathrm{ut}}_{s,k,0}
\big]^T$.
Based on the corrected angles, the satellite-side and UT-side steering vectors are reconstructed by the Kronecker product:
\begin{align}
\hat{\mathbf g}_{s,k}
&=
\mathbf a_{M_{\rm x}}
\left(
\sin \hat{\theta}^{\rm sat}_{s,k}
\cos \hat{\phi}^{\rm sat}_{s,k}
\right)
\otimes
\mathbf a_{M_{\rm y}}
\left(
\cos \hat{\theta}^{\rm sat}_{s,k}
\right), \\
\hat{\mathbf d}_{s,k,0}
&\!=\!
\mathbf a_{N_{\rm x^{\prime}}}\!\!
\left(
\sin \hat{\theta}^{\rm ut}_{s,k,0}
\cos \hat{\phi}^{\rm ut}_{s,k,0}
\right)
\!\otimes\!
\mathbf a_{N_{\rm y^{\prime}}}\!\!
\left(
\cos \hat{\theta}^{\rm ut}_{s,k,0}
\right).
\end{align}
Then, we apply a per-antenna phase correction with normalization to the reconstructed steering vectors as follows:
\begin{align}
\tilde{\mathbf{g}}_{s,k}&\textstyle=\frac{\hat{\mathbf{g}}_{s,k} \odot \exp \left(j \Delta \boldsymbol{\psi}_{s,k}^{g}\right)}{\left\|\hat{\mathbf{g}}_{s,k} \odot \exp \left(j \Delta \boldsymbol{\psi}_{s,k}^{g}\right)\right\|_{2}},
\tilde{\mathbf{d}}_{s,k,0}\textstyle=\frac{\hat{\mathbf{d}}_{s,k,0} \odot \exp \left(j \Delta \boldsymbol{\psi}_{s,k}^{d}\right)}{\left\|\hat{\mathbf{d}}_{s,k,0} \odot \exp \left(j \Delta \boldsymbol{\psi}_{s,k}^{d}\right)\right\|_{2}}.
\end{align}

\vspace{-2mm}
\subsection{Training Objective and Distillation Loss Design}
\vspace{-1mm}
\label{subsec:loss_design}
The student NN is trained using a hybrid objective that combines a task-driven WSR loss with hybrid KD losses. The WSR loss directly optimizes the final task performance, whereas the distillation losses transfer structured knowledge from the teacher at both the precoder and steering-vector levels, as in \figref{fig:kd_design_framework}. During student training, the teacher NN remains fixed and serves only to provide supervision targets.
\subsubsection{\textbf{WSR Optimization}}
At each training iteration, the student NNs output the precoders
$\{\mathbf p_{s,k}^{\rm stu}\}_{\forall s,k}$, based on which the achievable rate
is evaluated using the instantaneous CSI. The task loss is defined as the
negative sample-averaged WSR:
\begin{align}
\!\!\!\!\mathcal{L}_{\rm obj}^{\rm wsr}
\textstyle\!=\!
-\frac{1}{N_{\rm sam}}\!\!
\sum_{n=1}^{N_{\rm sam}}\!
\sum_{s=1}^{S}\!
\sum_{k=1}^{K}\!\!
a_{s,k}R_{s,k}^{(n)}\Big(\{\mathbf p_{s,k}^{\rm stu}\}_{\forall s,k}^{(n)}\Big),
\end{align}
where $N_{\rm sam}$ is the number of training samples. Minimizing
$\mathcal{L}_{\rm obj}^{\rm wsr}$ is equivalent to maximizing the system WSR.
\begin{algorithm}[t]
  \caption{Decentralized Student NN with KD}
  \label{alg:student_kd}
  \begin{algorithmic}[1]
    \STATE \textbf{Input:} $\tT$, $\{\tS_s\}_{\forall s}$, $\mathcal{N}_{\rm tea}$, and $\{\alpha_P,\alpha_d,\alpha_g\}$.
    \STATE \textbf{Freeze the teacher NN $\mathcal{N}_{\rm tea}$ for supervision.}
    \STATE \textbf{Training:}
    $  \mathcal{N}_{\rm stu\text{-}kd}^{(0)}
    \!\!=\!\!
   \big\{\{N_{\rm STE}^{\rm stu\text{-}kd}\!,N_{\rm DAA}^{\rm stu\text{-}kd}\!,N_{\rm TEN}^{\rm stu\text{-}kd}\!,N_{\rm PD}^{\rm stu\text{-}kd}\}$\\
   \quad \quad \quad \quad \quad $\cup\{N_{\rm APRP}^{\rm stu\text{-}kd}\big\}^{(0)}$ and $n=0$.
    \STATE \textbf{for} $n<N_{\rm ep}$ \textbf{do}
      \STATE \quad \textbf{forward:}
      \STATE \quad \textbf{for} each satellite $s\in\mathcal{S}$ \textbf{do}
        \STATE \quad\quad 
         $
       \tF_s^{\rm stu} = \mathrm{Token}(\tS_s)
        $. 
        \STATE \quad\quad
        $
       \tX_s^{\rm stu\text{-}kd} = N_{\rm STE}^{\rm stu\text{-}kd}(\tF_s^{\rm stu})
        $.
        \STATE \quad\quad 
        $
       \tH_s^{\rm stu\text{-}kd} = N_{\rm DAA}^{\rm stu\text{-}kd}(\tX_s^{\rm stu\text{-}kd})
        $.
        \STATE \quad\quad  ${\bar{\mathbf{H}}^{\rm stu\text{-}kd}_{s}}=\frac{1}{S}\sum_{s^{\prime}=1}^{S}\tH^{\rm stu\text{-}kd}_{s[s^{\prime},:,:]}$.
        \STATE \quad\quad  
        $
       {\mathbf{E}^{\rm stu\text{-}kd}_{s}} = N_{\rm TEN}^{\rm stu\text{-}kd}({\bar{\mathbf{H}}^{\rm stu\text{-}kd}_{s}})
        $.
        \STATE \quad\quad $ \{\!\Delta\hat{\boldsymbol{\Psi}}_{s,k},\!\Delta{\boldsymbol{\psi}}_{s,k}^{ g},\!\Delta{\boldsymbol{\psi}}_{s,k}^{d}\}_{\forall k}
        \!\!=\!\!
        N_{\rm APRP}^{\rm stu\text{-}kd}(\bar{\mathbf{H}}^{\rm stu-kd}_{s})$.
        \STATE \quad\quad 
        $\{\tilde{\mathbf d}_{s,k,0},\!
        \tilde{\mathbf g}_{s,k}\}_{\forall k}
        \!=\!\!
        N_{\rm KSVC}
        \big(\!
        \{\!\Delta\!\hat{\boldsymbol{\Psi}}_{s,k},\!\Delta\!{\boldsymbol{\psi}}_{s,k}^{ g},\!\Delta\!{\boldsymbol{\psi}}_{s,k}^{d}\!\}_{\forall k}
        \!\big)$.
       \STATE \quad\quad  
        $
       {\mathbf{C}^{\rm stu\text{-}kd}_{s}} = N_{\rm PD}^{\rm stu\text{-}kd}({\mathbf{E}^{\rm stu\text{-}kd}_{s}})
        $.
        \STATE \quad\quad 
        $\{\mathbf b_{s,k}^{\rm stu\text{-}kd},\mathbf p_{s,k}^{\rm stu\text{-}kd}\}_{\forall k}
        \!\!=\!\!  N_{\rm PR}^{\rm stu}
        \big(\!
        {\mathbf C}_s^{\rm stu\text{-}kd}\!\!\!,
        \{\tilde{\mathbf d}_{s,k,0},
        \tilde{\mathbf g}_{s,k}\}_{\forall k}\!
        \big)$.
      \STATE \quad \textbf{end for}
      \STATE \quad Compute the overall loss $\mathcal{L}_{\rm stu\text{-}kd}$ according to
\eqref{eq:student_total_loss}, with $\tT$ fed into $\mathcal{N}_{\rm tea}$
and $\{\tS_s\}_{\forall s}$ fed into $\mathcal{N}_{\rm stu\text{-}kd}^{(n)}$.
      \STATE \quad \textbf{backward:}
     Update $\mathcal{N}_{\rm stu\text{-}kd}^{(n)}$ via $\mathcal{L}_{\rm stu\text{-}kd}$.
      \STATE \quad Set $n=n+1$.
    \STATE \textbf{end for}
    \STATE Store the learned variables of $\mathcal{N}_{\rm stu\text{-}kd}$.
    \STATE \textbf{Inference:} For $\forall s$, use the partial noisy input $\tS_s$
    and the trained $\mathcal{N}_{\rm stu\text{-}kd}$ to obtain
    $\{\mathbf p_{s,k}^{\rm stu\text{-}kd},\mathbf b_{s,k}^{\rm stu\text{-}kd}\}_{\forall k}$ in parallel.
  \end{algorithmic}
\end{algorithm}
\subsubsection{\textbf{Precoder Distillation}}
Let
$\mathbf P_s^{\rm tea}=[\mathbf p_{s,1}^{\rm tea},\ldots,\mathbf p_{s,K}^{\rm tea}]$
and
$\mathbf P_s^{\rm stu}=[\mathbf p_{s,1}^{\rm stu},\ldots,\mathbf p_{s,K}^{\rm stu}]$
denote the teacher and student precoder matrices of satellite $s$, respectively. For complex-valued vectors or matrices, we adopt the cosine similarity \cite{wang2026}:
\begin{align}
\ell_{\rm cos}(\hat{\mathbf{X}}, \mathbf{X})\textstyle=1-\frac{\left|\operatorname{vec}(\hat{\mathbf{X}})^{H} \operatorname{vec}(\mathbf{X})\right|}{\|\operatorname{vec}(\hat{\mathbf{X}})\|\|\operatorname{vec}(\mathbf{X})\|+\epsilon},
\label{eq:cos}
\end{align}
 $\epsilon\!>\!0$ is for numerical stability. The precoder-level KD loss is:
\begin{align}
\!\!\mathcal{L}_{\rm kd}^{P}
\textstyle\!\!=\!\!
\frac{1}{N_{\rm sam}S}
\sum_{n=1}^{N_{\rm sam}}
\sum_{s=1}^{S}
\ell_{\rm cos}
\left(
 (\mathbf{P}_s^{\rm stu})^{(n)},
{(\mathbf P_s^{\rm tea}})^{(n)}
\right).
\label{eq:kd_p_loss}
\end{align}
\subsubsection{\textbf{Steering Vector Distillation}}
Based on the corrected angles and phase, the student reconstructs the steering vectors
$\tilde{\mathbf d}_{s,k,0}$ and $\tilde{\mathbf g}_{s,k}$. The corresponding teacher-side steering vectors are denoted by
$\mathbf d_{s,k,0}$ and $\mathbf g_{s,k}$. The steering-vector KD losses are formulated as follows:
\begin{align}
\mathcal{L}_{\rm kd}^{d}
&\textstyle\!=\!
\frac{1}{N_{\rm sam}SK}\!
\sum_{n=1}^{N_{\rm sam}}\!
\sum_{s=1}^{S}\!
\sum_{k=1}^{K}\!
\ell_{\rm cos}\!
\left(\!
\tilde{\mathbf d}_{s,k,0}^{(n)},
\mathbf d_{s,k,0}^{(n)}\!
\right),\\
\mathcal{L}_{\rm kd}^{g}
&\textstyle=
\frac{1}{N_{\rm sam}SK}
\sum_{n=1}^{N_{\rm sam}}
\sum_{s=1}^{S}
\sum_{k=1}^{K}
\ell_{\rm cos}
\left(
\tilde{\mathbf g}_{s,k}^{(n)},
\mathbf g_{s,k}^{(n)}
\right).
\end{align}
These two losses align the geometry-consistent steering vectors reconstructed by the student with those generated from the teacher's clean angle information.

Finally, the overall student loss is written as:
\begin{align}
\mathcal{L}_{\rm stu\text{-}kd}
\!=\!
\mathcal{L}_{\rm obj}^{\rm wsr}
\!+\!
\alpha_{P}\mathcal{L}_{\rm kd}^{P}
\!+\!
\alpha_{d}\mathcal{L}_{\rm kd}^{d}
\!+\!
\alpha_{g}\mathcal{L}_{\rm kd}^{g},
\label{eq:student_total_loss}
\end{align}
where $\alpha_{P}$, $\alpha_{d}$, and $\alpha_{g}$ control the relative importance of different KD losses.
The entire student NN using KD is summarized in \algref{alg:student_kd}.

The dataset comprises 10,000 samples, with 7,000 used for training, 2,000 for validation, and an additional 1,000 for testing. Each sample randomly selects the center of the cooperative transmission region within the global coverage of the constellation, generates UTs, and determines the cooperating satellites. \textcolor{black}{It further includes data corresponding to five transmit power levels, namely $P_s^{\rm sat} \in  \{-10, -5, 0, 5, 10\}\, \text{dBW}$, which are randomly selected during training to enable the NN to operate under various link budgets resulting from various transceiver configurations \cite{Xiang2024,3GPPTR38.821,wang2025MSMS}.}

\vspace{-2mm}
\section{Numerical Results}\label{simulation}
\vspace{-1mm}
We employ the QuaDRiGa channel simulator to generate the propagation scenarios and radio channel parameters. Specifically, the channel parameters are obtained using the QuaDRiGa\_NTN\_Urban\_LOS\_scenario \cite{Burkhardt2014}. This simulator, with appropriate parameter calibration, is consistent with the channel model considered in this work as well as the Third Generation Partnership Project specifications. Monte Carlo simulations are conducted by randomly selecting a point within the constellation coverage as the center, defining a circular region with the specified radius, and choosing the \(S\) nearest satellites to that center for multi-satellite cooperative transmission. In addition, all satellites are assumed to have identical power constraints, i.e., $P_s^{\rm sat} = P^{\rm sat}, \forall s$.
The remaining simulation parameters are summarized in \tabref{param}.
	\begin{table}[!t]
    \centering
    \caption{Multi-Satellite System Parameters \cite{FCC_SpaceX_Gen2_2021,3GPPTR38.821,Xiang2024,wang2026}}
    \label{param}
    \resizebox{0.7\columnwidth}{!}{%
    \begin{tabular}{cc}
    \toprule
    \textbf{Parameter}  &  \textbf{Value}  \\
    \midrule
    Constellation Type & Walker-Delta \\
    Satellite Orbit Altitude & $600 \ \rm{km}$ \\
    Orbital Planes & 28 \\
    Satellites Per Plane & 60 \\
    Inclination  & $53^{\circ}$\\
    Multi-Satellite Coverage Radius & $800 \ \rm{km}$ \\
    Number of Cooperating Satellites & 2,3,4\\
    \midrule
    Number of UTs & 12,18,24,30,36 \\ 
    Distribution of UTs & Uniform\\
    Velocity of UTs & $5 \ \rm{km/h}$ \\
    \midrule
    Carrier Frequency & $2\ \rm{GHz}$\\
    Subcarrier Spacing & $30 \ \rm {kHz}$ \\
    System Bandwidth (DL) & $20\ \rm{MHz}$\\
    \midrule
    Transmit Antenna Number $M$ & $64$\\
    Receive Antenna Number $N$ & $4$\\
    Per-Antenna Gain $ G_{\rm T}$, $G_{\rm R}$ & $6 \ \rm{dBi}$, $ 0\ \rm{dBi}$ \\
    Noise Figure $F$ & $7 \ \rm {dB}$\\
    Noise Temperature $T$ & $290 \ \rm {K}$\\
    \bottomrule
    \end{tabular}%
    }
\end{table}
\subsection{Performance Comparison}
This subsection evaluates the effectiveness of the proposed hybrid KD framework. The following schemes are compared.
\begin{itemize}
    \item \textbf{Sep-WMMSE (clean)}: The separate WMMSE optimization independently
    performed at each satellite, without joint optimization of inter-satellite
    interference \cite{Christensen2008, Shi2011}.

\item \textbf{Cen-WMMSE (clean/noisy)}: A centralized iterative WMMSE optimization algorithm with global CSI and cooperative multi-satellite transmission \cite{Xiang2024,cao2026deep}.

  \item \textbf{Cen-Teacher (clean/noisy)}: The centralized teacher NN, whose architecture is described in \secref{teacher_network}, is trained with global clean sCSI observations. During evaluation, its performance is tested under both clean and noisy global sCSI inputs, denoted by Cen-Teacher (clean) and Cen-Teacher (noisy), respectively.

   \item \textbf{Dec-Student (noisy)}: The decentralized student NN is trained with partial and noisy inputs, without employing KD, as described in \secref{student_network}.

  \item \textbf{Dec-Student-PKD (noisy)}: Based on Dec-Student, we incorporate precoder-level KD from the teacher.

    \item \textbf{Dec-Student-HKD (noisy)}: The proposed decentralized student
    NN trained with the WSR objective and hybrid KD (HKD),
    consisting of precoder-level and steering-vector-level distillation, with angle correction and steering-vector phase correction, as in \secref{Knowledge_Distillation_Design}.
\end{itemize}

\begin{figure}[!t]
    \centering  
\includegraphics[width=0.78\linewidth,trim=3.2cm 9.2cm 3cm 9.2cm,clip]{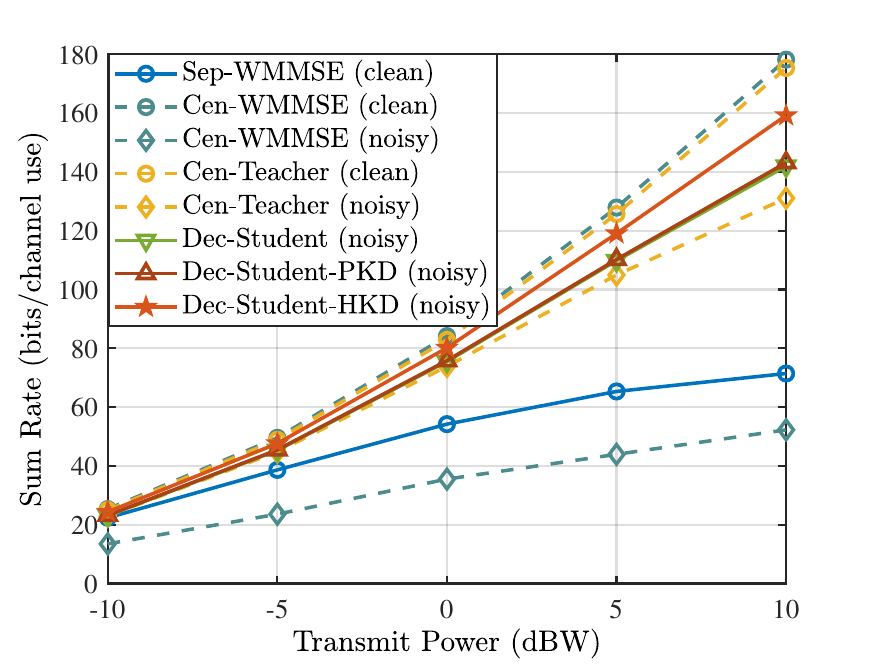}    
    \caption{Multi-satellite sum rate vs ${P}^{\rm sat}$.}   \label{1_wsr_teacher_student_kd}
    \vspace{-4mm}
\end{figure}

\figref{1_wsr_teacher_student_kd} compares the multi-satellite sum rate versus the satellite transmit power under $S=3$ and $K=24$. It is observed that Cen-Teacher (clean) closely approaches Cen-WMMSE (clean) over the entire transmit-power range, verifying that the large teacher NN can effectively learn the centralized global precoding policy from clean observations. For decentralized inference, Dec-Student exhibits an evident performance loss due to its limited observation scope and noisy angle inputs, but it still outperforms Sep-WMMSE (clean), which confirms the necessity of satellite cooperation transmission.
The proposed Dec-Student-HKD consistently outperforms Dec-Student and Dec-Student-PKD with an increasingly noticeable gain at larger $P^{\rm sat}$. This demonstrates the effectiveness of the proposed hybrid KD, where the joint supervision on steering vectors and precoders provides effective guidance.
It is also observed that Cen-WMMSE (noisy) and Cen-Teacher (noisy) suffer from severe performance degradation compared with their clean counterparts. This indicates that directly applying centralized schemes to noisy partial sCSI is highly sensitive to observation errors. By contrast, Dec-Student-HKD is explicitly trained for noisy partial sCSI and achieves a much higher sum rate, confirming the necessity and robustness of the proposed KD framework.
\tabref{tab:cs_precoder} validates the effectiveness of the proposed hybrid KD scheme in precoders. Compared with Dec-Student, Dec-Student-HKD consistently achieves higher precoder cosine similarity across all transmit power levels. Specifically, the proposed distillation scheme brings a stable cosine-similarity gain of about 7\%, indicating that hybrid KD effectively guides the student NN to recover precoders with better directional alignment. This improvement demonstrates that the transferred steering-vector-level and precoder-level knowledge enhances the effectiveness and robustness of decentralized precoding under partial noisy sCSI.
\begin{table}[!tbp]
\centering
\caption{Cosine similarity comparison of the precoders.}
\label{tab:cs_precoder}
\renewcommand{\arraystretch}{1}
\setlength{\tabcolsep}{3pt}
\resizebox{\columnwidth}{!}{%
\begin{tabular}{@{}lccccc@{}}
\toprule
\textbf{Schemes}
& \multicolumn{5}{c}{\textbf{Transmit power (dBW)}} \\
\cmidrule(l){2-6}
& \(\mathbf{-10}\)
& \(\mathbf{-5}\)
& \(\mathbf{0}\)
& \(\mathbf{5}\)
& \(\mathbf{10}\)
\\
\midrule
Dec-Student (noisy)
& 0.8021
& 0.7938
& 0.7761
& 0.7519
& 0.7239
\\
Dec-Student-HKD (noisy)
& 0.8588
& 0.8471
& 0.8295
& 0.8087
& 0.7837
\\
\midrule
{Cosine-similarity gain}
& \({+0.0567}\)
& \({+0.0533}\)
& \({+0.0534}\)
& \({+0.0568}\)
& \({+0.0599}\)
\\
\textbf{Gain percentage}
& \(\mathbf{+7.07 \,\%}\)
& \(\mathbf{+6.71 \,\%}\)
& \(\mathbf{+6.88 \,\%}\)
& \(\mathbf{+7.55 \,\%}\)
& \(\mathbf{+8.27 \,\%}\)
\\
\bottomrule
\end{tabular}%
}
\end{table}

\begin{figure}[!t]
  \centering  
\includegraphics[width=0.75\linewidth,trim=0.1cm 8.5cm 0.1cm 7.8cm,clip]{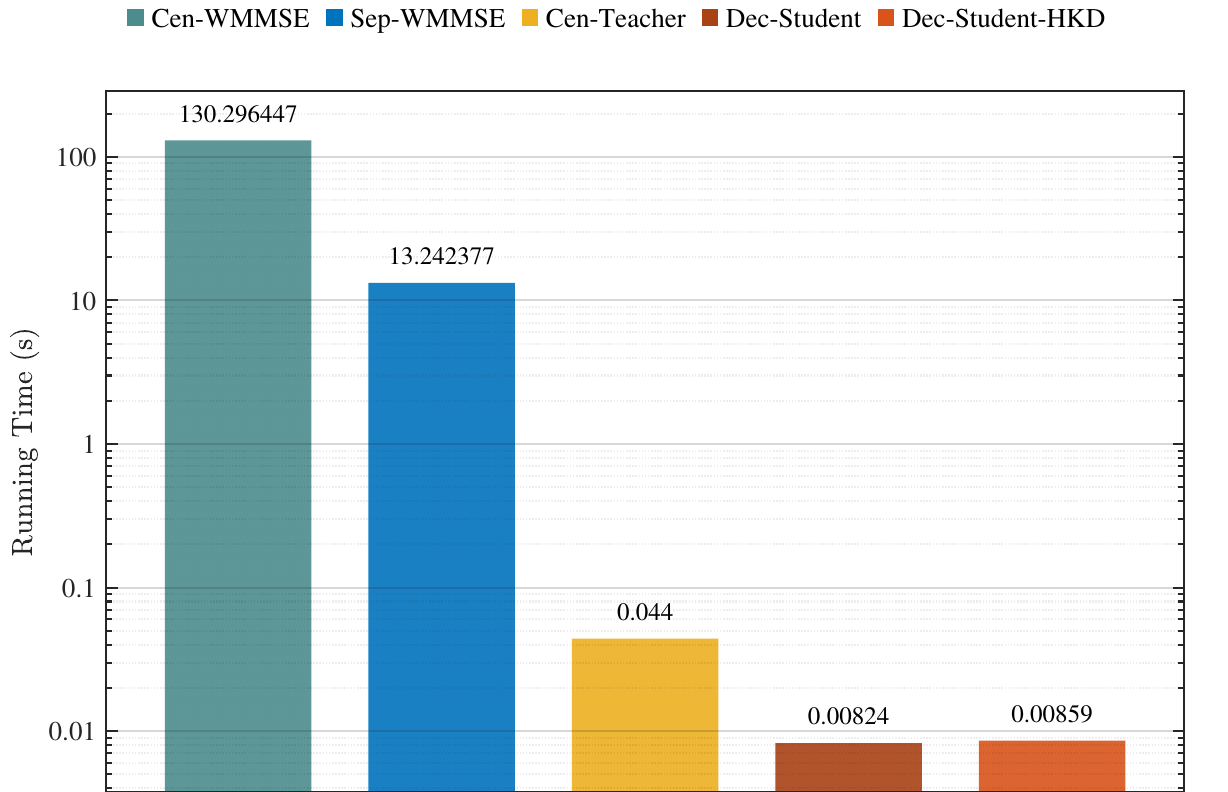} 
    \caption{Running time comparison.}
  \label{5_running_time_kd}
  \vspace{-5mm}
\end{figure}
\begin{table}[!t]
\centering
\caption{Computational and inference complexity order comparison.}
\label{tab:complexity}
\renewcommand{\arraystretch}{1}
\setlength{\tabcolsep}{1pt}
\scriptsize
\begin{tabularx}{\columnwidth}{@{}
    >{\raggedright\arraybackslash}p{0.22\columnwidth}
    >{\centering\arraybackslash}X
    >{\centering\arraybackslash}p{0.15\columnwidth}
    @{}}
\toprule
\textbf{Scheme}
&
\textbf{Complexity order}
&
\textbf{Mode}
\\
\midrule
Sep-WMMSE
&
\(\mathcal{O}\!\left(
I_{\max}^{\mathrm{out}}
I_{\max}^{\mathrm{in}}
K M^{3}
\right)\)
&
Parallel
\\
\midrule

Cen-WMMSE
&
\(\mathcal{O}\!\left(
I_{\max}^{\mathrm{out}}
I_{\max}^{\mathrm{in}}
S K M^{3}
\right)\)
&
Centralized
\\
\midrule
Cen-Teacher
&
\(
\begin{aligned}
\mathcal{O}\Big(
& S K L T^{2}D_{\mathrm{h}}
 +L_{\mathrm{axis}}S K(S+K)D_{\mathrm{h}}
\\[-3pt]
& {}+L_{\mathrm{ten}}S K D_{\mathrm{ten}}^{2}
 +S K M^{3}
\Big)
\end{aligned}
\)
&	
Centralized
\\
\midrule
Dec-Student
&
\(
\begin{aligned}
\mathcal{O}\Big(
& S K T^{2}D_{\mathrm{h}}
 +L_{\mathrm{axis}}S K(S+K)D_{\mathrm{h}}
\\[-3pt]
& {}+L_{\mathrm{ten}}S K D_{\mathrm{ten}}^{2}
 +K M^{3}
\Big)
\end{aligned}
\)
&
Parallel
\\
\midrule
Dec-Student-HKD
&
\(
\begin{aligned}
\mathcal{O}\Big(
& S K T^{2}D_{\mathrm{h}}
 +L_{\mathrm{axis}}S K(S+K)D_{\mathrm{h}}
\\[-3pt]
& {}+L_{\mathrm{ten}}S K D_{\mathrm{ten}}^{2}
 +K M^{3}+K(M+N)
\Big)
\end{aligned}
\)
&
Parallel
\\
\bottomrule
\end{tabularx}
\end{table}
\begin{figure}[!t]
    \centering  
\includegraphics[width=1\linewidth,trim=0.01cm 0.01cm 0.01cm 0.01cm,clip]{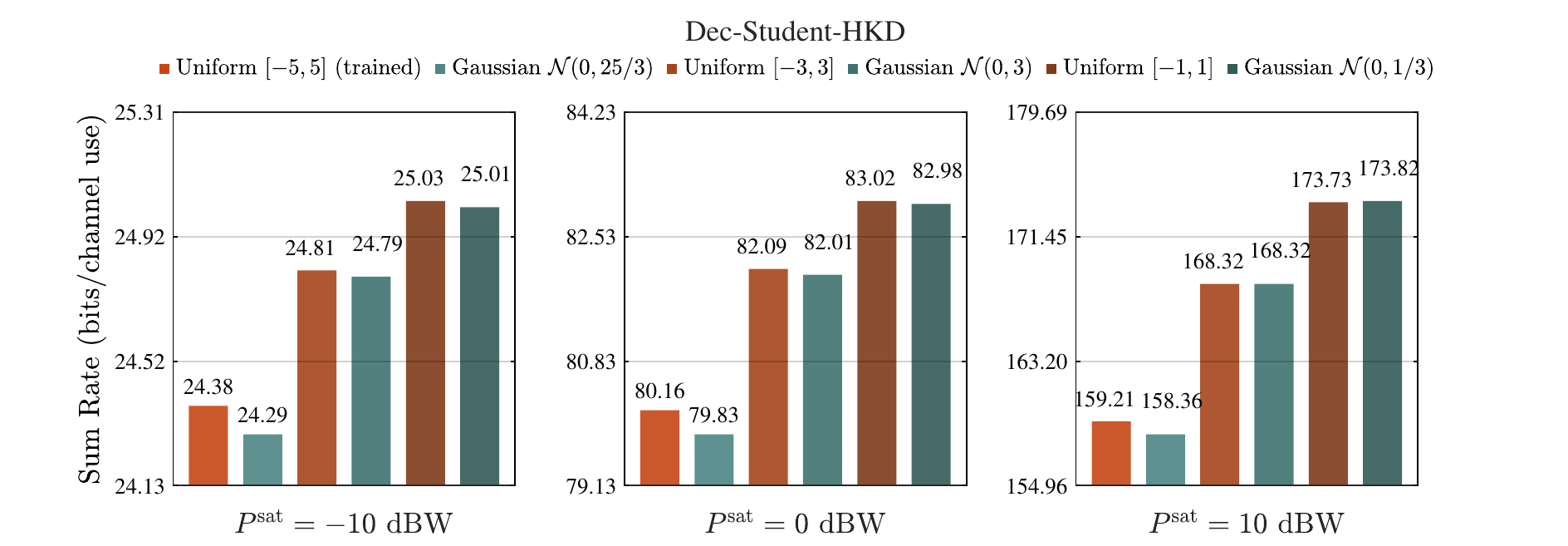} \\   
\includegraphics[width=1\linewidth,trim=0.01cm 0.01cm 0.01cm 0.01cm,clip]{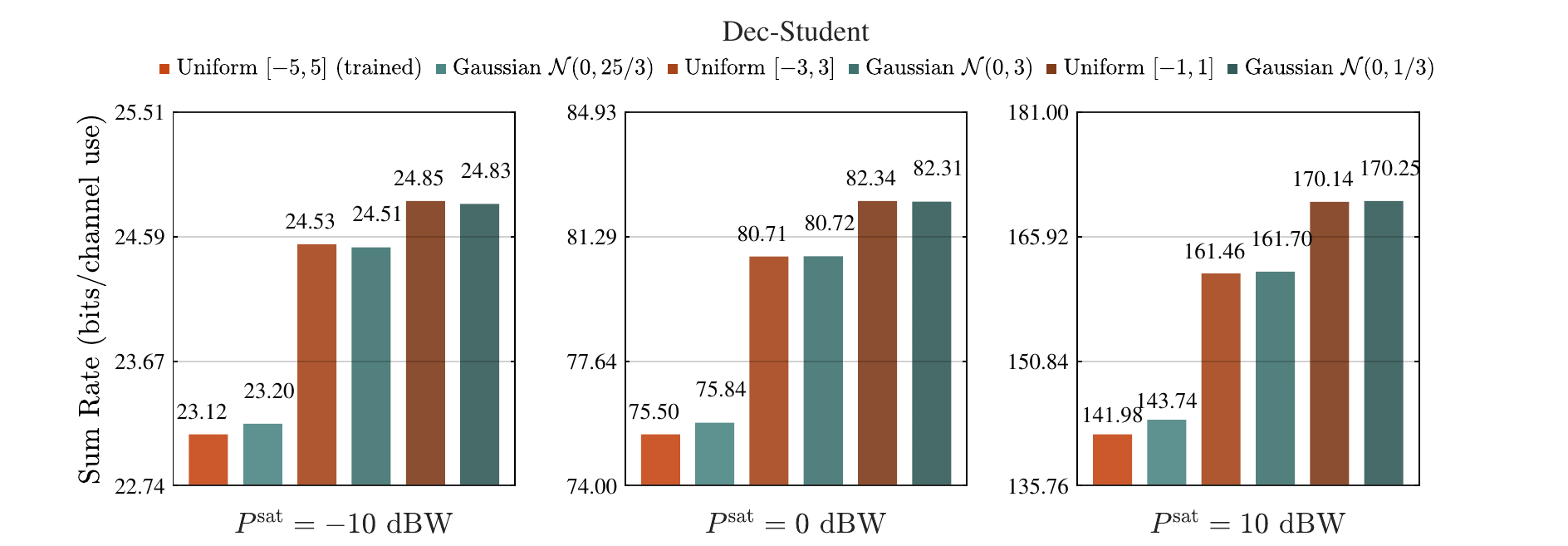} \\
\includegraphics[width=1\linewidth,trim=0.01cm 0.01cm 0.01cm 0.01cm,clip]{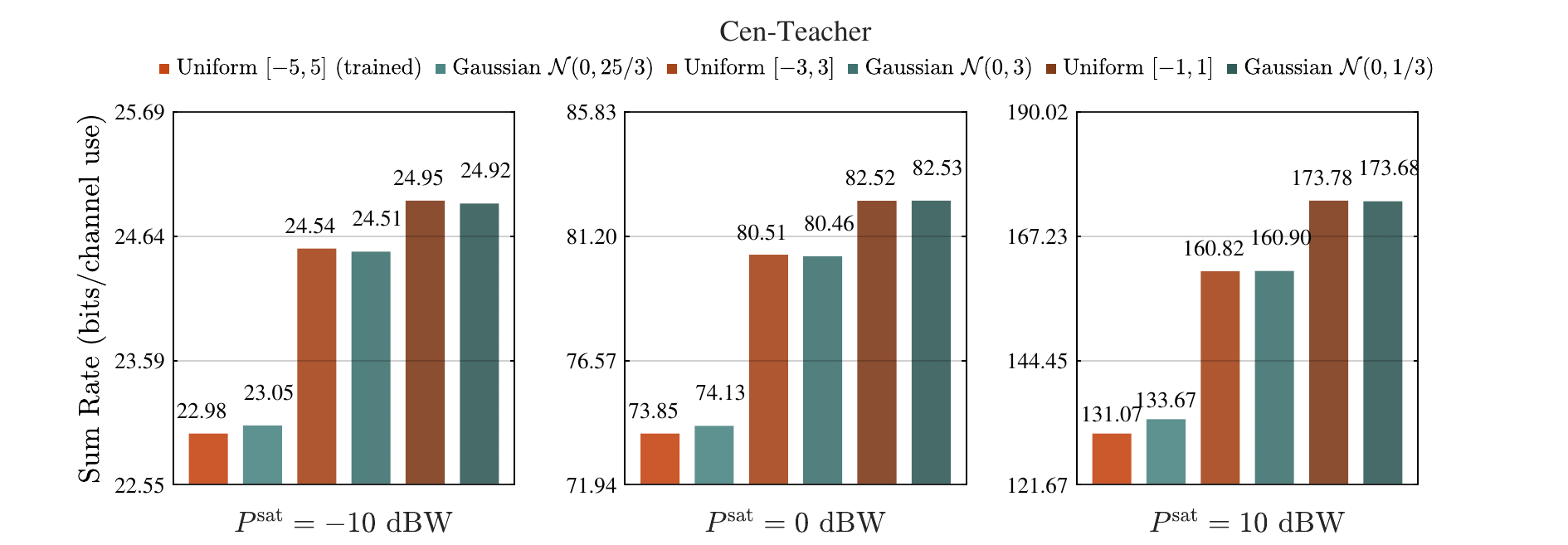}      
    \caption{Different angle error distribution generalization.}
\label{2_noisy_uniform_gaussian}
  \vspace{-5mm}
\end{figure}

As summarized in \tabref{tab:complexity}, the optimization-based schemes require iterative WMMSE updates, leading to the iteration number \(I_{\max}^{\mathrm{out}}I_{\max}^{\mathrm{in}}\). In contrast, learning-based schemes perform only one neural forward pass followed by closed-form precoder recovery, thereby avoiding iterative optimization and greatly reducing online inference complexity. Moreover, the decentralized student NNs are lighter than the centralized teacher NN by replacing the \(L\)-layer Transformer with a single-layer structure. Although Dec-Student-HKD introduces an additional low-order term \(K(M+N)\) for steering-vector recomputation, the dominant cost remains approximately \(K M^3\). Therefore, all student schemes have nearly identical inference complexity.
\figref{5_running_time_kd} compares the running time of different schemes on a laptop platform with an Intel(R) Core(TM) Ultra 9 185H CPU @ 2.30 GHz. Due to iterative optimization, Cen-WMMSE and Sep-WMMSE incur much higher latency than the learning-based schemes, which only require non-iterative inference after training. Specifically, Cen-Teacher achieves a running time of 0.044 s, while the lightweight decentralized students further reduce it to below 0.01 s. This indicates that Dec-Student-HKD supports efficient real-time inference while preserving the benefits of structured knowledge transfer. The measured time is implementation- and hardware-dependent and should be viewed as indicative.

\vspace{-4mm}
\subsection{Robustness to Angle Error Distributions}
\vspace{-1mm}
\figref{2_noisy_uniform_gaussian} evaluates the robustness of the proposed hybrid KD framework under different angle-error distributions. Dec-Student-HKD consistently outperforms Dec-Student across all tested settings. More importantly, under large angle perturbations, Dec-Student-HKD substantially outperforms Cen-Teacher (noisy). This result indicates that directly feeding noisy observations into the centralized teacher makes it highly sensitive to angle errors, whereas the proposed distillation framework explicitly equips the decentralized student with angle- and phase-correction capabilities. As the angle error decreases, the performance of Dec-Student-HKD becomes comparable to, and occasionally slightly better than, that of Cen-Teacher (noisy).  Moreover, the proposed method remains stable under both uniform and Gaussian angle errors. Although the two error models have different distributional shapes, Dec-Student-HKD exhibits similar performance trends when their means and variances are matched, while consistently outperforming Dec-Student. These results suggest that the proposed distillation framework does not overfit to a specific angle-error distribution. Instead, it enables the student to acquire distribution-robust angle-correction knowledge for practical decentralized multi-satellite transmission.
\begin{figure}[!t]
    \centering  
\includegraphics[width=1\linewidth,trim=0.01cm 0.01cm 0.01cm 0.01cm,clip]{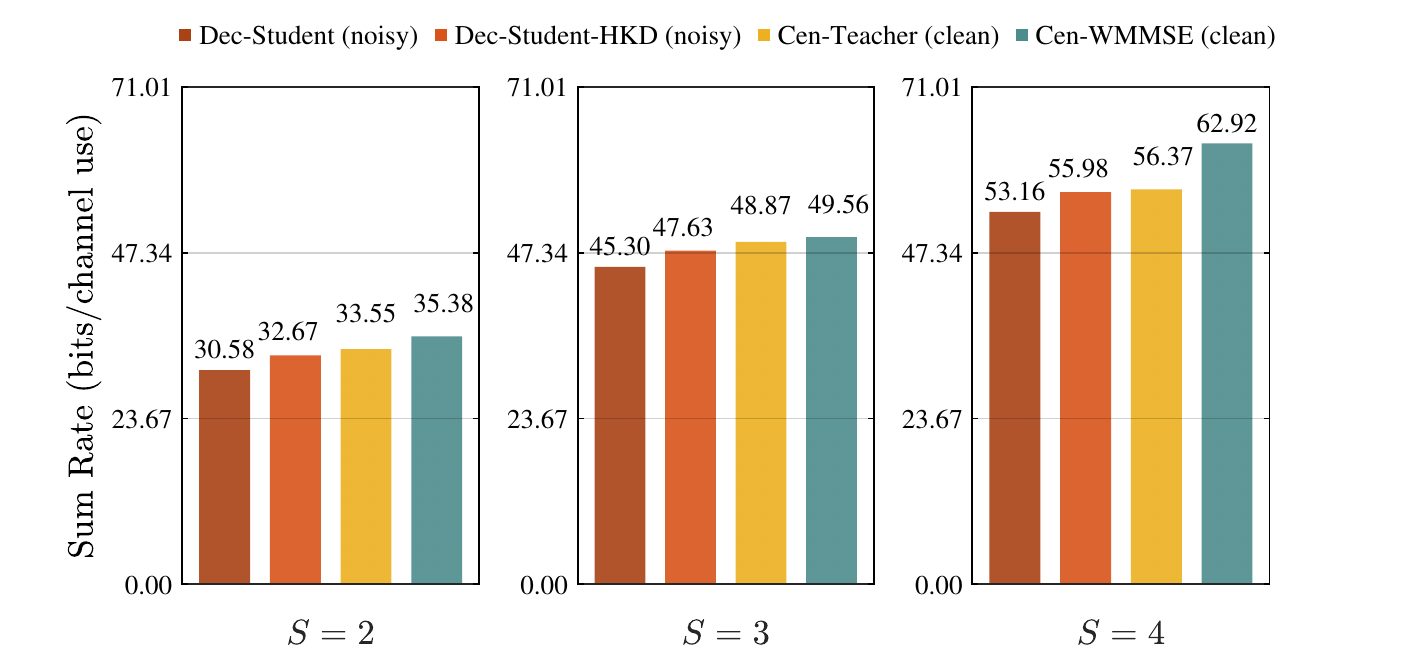}    
    \caption{Multi-satellite sum rate vs satellite number with $K\!=\!24$ and
    $P^{\rm sat}\!=\!-10\, \rm{dBW}$, where NN schemes are all trained at $S\!=\!3$.}
    \label{3_sat_generalize}
    \vspace{-4mm}
\end{figure}
\begin{figure}[!t]
    \centering  
\includegraphics[width=0.7\linewidth,trim=3.8cm 9.2cm 4cm 9.5cm,clip]{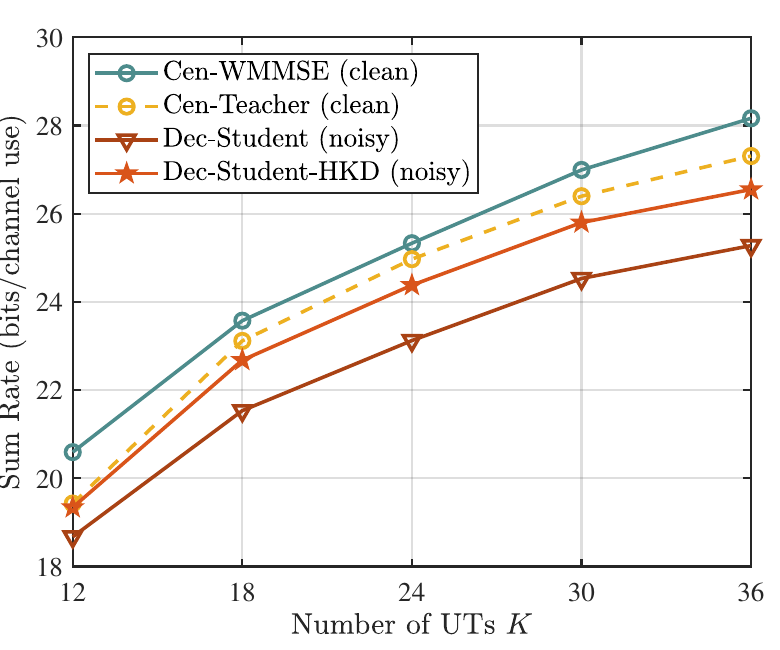}    
    \caption{Multi-satellite sum rate vs UT number with $S=3$ and
    $P^{\rm sat}=-10\, \rm{dBW}$, where NN schemes are all trained at $K=24$.}
  \label{4_ut_generalize}
    \vspace{-5mm}
\end{figure}
\vspace{-3mm}
\subsection{Scalability to Network Scenario Size}
\vspace{-1mm}
\figref{3_sat_generalize} and \figref{4_ut_generalize} evaluate the scalability of the proposed method with respect to the number of satellites and UTs, respectively. All learning-based schemes are trained under a fixed configuration with $S=3$ and $K=24$, and are directly tested under different network sizes without retraining.
In \figref{3_sat_generalize}, the sum rate increases with the number of cooperative satellites. Among the decentralized schemes, Dec-Student-HKD remains consistently closer to Cen-Teacher and Cen-WMMSE than Dec-Student, demonstrating the robustness and transferability of hybrid KD across different satellite scales. In \figref{4_ut_generalize}, all schemes achieve higher sum rates as the number of UTs increases. The proposed Dec-Student-HKD maintains a stable advantage over the Dec-Student across the whole range of $K$, indicating its scalability to different user loads.
\vspace{-1mm}
\section{Conclusion}
\label{conclusion}
\vspace{-1mm}
This paper has investigated robust decentralized transmission for cooperative multi-satellite massive MIMO systems with partial and imperfect sCSI. A teacher-student KD framework for multi-satellite cooperative transmission is proposed, where a centralized teacher learns high-performance precoders from clean global sCSI, and lightweight decentralized students perform onboard inference using only noisy partial sCSI and limited ISL-exchanged information. To reduce the teacher-student performance gap, we design an angle- and phase-correction-based hybrid KD mechanism, which calibrates angle errors and the resulting phase errors, reconstructs steering vectors, and performs hybrid distillation. 
Simulation results demonstrated that the proposed scheme significantly improves the sum-rate performance of decentralized student NNs with negligible additional inference overhead. 

	
	%

	
	
\vspace{-2mm}
\appendices

	\ifCLASSOPTIONcaptionsoff
	\newpage
	\fi
	\bibliographystyle{IEEEtran}
	\bibliography{IEEEfull}

\end{document}